\documentclass[preprint,12pt,numbers]{elsarticle}

\usepackage[a4paper,margin=2cm]{geometry}

\usepackage{graphicx}
\usepackage{booktabs}
\usepackage{multirow}
\usepackage{amsmath,amssymb}
\usepackage{siunitx}
\usepackage{subcaption}
\usepackage{enumitem}
\usepackage{tabularx}
\usepackage{float}

\usepackage{hyperref}
\hypersetup{
    hidelinks,
    pdftitle={A Multi-Level Preprocessing and Modelling Framework for Spectral Imaging of Microplastics},
    pdfauthor={Zina-Sabrina Duma, Tenzin Tsering, Sara Heikkinen, Tuomo Soininen, Tuomas Sihvonen, Arto Koistinen, Satu-Pia Reinikainen},
    pdfsubject={FT-IR spectral imaging and microplastic identification},
    pdfkeywords={microplastics, FT-IR spectral imaging, preprocessing, clustering, spectral matching, classification}
}

\usepackage[nameinlink,noabbrev]{cleveref}

\setlist[itemize]{noitemsep,topsep=2pt}

\begin{document}

\begin{frontmatter}

\title{A Multi-Level Preprocessing and Modelling Framework for Spectral Imaging of Microplastics}

\author[1]{Zina-Sabrina Duma\corref{cor1}}
\ead{Zina-Sabrina.Duma@lut.fi}
\cortext[cor1]{Corresponding author}

\author[2,3]{Tenzin Tsering}
\author[1]{Sara Heikkinen}
\author[2]{Tuomo Soininen}
\author[1]{Tuomas Sihvonen}
\author[2]{Arto Koistinen}
\author[1]{Satu-Pia Reinikainen}

\address[1]{Department of Computational Engineering, School of Engineering Science, LUT University, Lappeenranta, Finland}
\address[2]{Department of Technical Physics, Faculty of Science, Forestry and Technology, University of Eastern Finland, Kuopio, Finland}
\address[3]{Department of Environmental and Biological Sciences, Faculty of Science, Forestry and Technology, University of Eastern Finland, Kuopio, Finland}

\begin{abstract}
Spectral imaging provides chemically specific and spatially resolved analysis of microplastics, but its routine application is hindered by large data volumes, acquisition artefacts, spectral variability, and misidentification of polymers due to alike spectra. This study proposes a multi-level preprocessing and modelling framework for FT-IR spectral imaging of microplastics that integrates image-level, tile-level, and spectral-level corrections with scalable identification strategies. 

Image-level variation associated with changing acquisition conditions was done with latent variable selection, while a background-based tile correction reduced illumination-related artefacts. Spectral preprocessing combined baseline correction, smoothing, derivative calculation, normalization, and wavelength selection, and only particle spectra were retained for further analysis to improve computational efficiency. For scalable identification, clustering was applied to particle spectra and spectral library matching was performed on cluster centroids instead of individual pixels. Among twelve evaluated matching strategies, a sign-invariant derivative-based cosine similarity method achieved perfect classification accuracy for polystyrene (PS), polyethylene terephthalate (PET), polyethylene (PE), and polypropylene (PP). The clustering-based workflow also produced more spatially coherent particle maps than direct software-based matching while substantially reducing processing time. The framework was evaluated for supervised classification-based MP indentification. These results show that multi-level correction combined with cluster-centroid spectral matching improves the robustness, efficiency, and interpretability of spectral-imaging-based microplastic identification.
\end{abstract}

\begin{keyword}
Microplastic identification \sep FT-IR spectral imaging \sep multi-level preprocessing \sep clustering \sep spectral matching \sep supervised classification
\end{keyword}

\end{frontmatter}

\section{Introduction}

Microplastics (MPs) are now widely recognized as emerging and persistent contaminants in aquatic and terrestrial environments, motivating the development of reliable analytical workflows for detection, identification, and quantification \citep{Fakayode2024}. While visual inspection remains common for assessing particle abundance and shape, it is inherently subjective and susceptible to observer bias \citep{Valente2023}. Spectral techniques such as Fourier transform infra-red (FT-IR) and Raman spectroscopy provide chemically specific identification, and their imaging variants (FPA-FTIR, hyperspectral and multispectral imaging) extend this capability by capturing both spatial and spectral information, enabling simultaneous identification and particle-level quantification across heterogeneous samples \citep{DaSilva2020,Faltynkova2021}.

Despite this promise, routine and scalable spectral-imaging-based MP analysis remains challenging. Spectral image cubes are high-dimensional and data-intensive, often containing millions of pixel spectra, which stresses storage, preprocessing, and modelling pipelines \citep{Faltynkova2021}. Furthermore, experimental factors can strongly modify measured spectra: baseline drift, scattering effects, instrumental artefacts, and spatial non-uniformities in illumination and shadowing reduce the robustness of downstream identification \citep{DaSilva2020,Faltynkova2021}. In environmental samples, additional within-class variance arises due to non-uniform weathering, thickness and particle size effects, and contamination or fouling, which may suppress diagnostic peaks or introduce noise and spurious features \citep{Faltynkova2021,Yan2022,VallsConesa2023}. These effects can render straightforward library search unreliable unless both the reference database and matching strategy are carefully designed \citep{VallsConesa2023}.

A broad range of identification strategies has therefore been explored. Traditional approaches include spectral library search (sometimes enhanced by peak detection, thresholding, or similarity indices), as well as protocol-driven matching frameworks (e.g., based on smoothed/baseline-corrected peak patterns) \citep{Simon2018,Renner2019,Primpke2017,Primpke2018,Primpke2019,Schmidt2018}. Beyond direct library search, chemometric and machine learning (ML) methods such as partial least-squares regression with discriminant analysis (PLS-DA), soft independent modelling of class analogy (SIMCA), support vector machines (SVM), k-Nearest neighbors (KNN), random forests (RFs), and ensembles have been applied to FT-IR imaging spectra, often reporting improved average classification performance compared to library-based approaches \citep{Kedzierski2019,Hufnagl2019,Michel2020,DaSilva2020,Yan2022,Faltynkova2021}. More recently, deep learning has been proposed for automated recognition; however, the practical adoption of such models is frequently constrained by data requirements and by limited evidence on robustness to chemically and physically altered spectra (e.g., additives, weathering, pollutants, and thickness variations) \citep{Zhu2023}. In addition, several studies highlight that many reported ML results emphasize aggregate scores, while MP datasets can be imbalanced and failure modes may concentrate on specific classes or mixtures \citep{Yan2022}.

A particularly important failure mode for MP identification is \emph{class confusability} driven by underlying chemistry. For example, polyethylene (PE) and polypropylene (PP) exhibit high molecular similarity because both are dominated by C--H and C--C bonds, leading to substantial overlap in spectral fingerprints and persistent confusion even when modelling capacity is increased \citep{Zhu2023,VallsConesa2023}. More generally, classification is impeded by large within-class variance and similarity across spectral signatures for several MP types; this motivates methods that explicitly characterize uncertainty and misclassification risk rather than only reporting mean accuracy \citep{VallsConesa2023}. Practical measurement constraints also contribute: spectral resolution can influence peak separability and classification precision \citep{Nicolau2024}, and poor reflectance from dark particles (e.g., grey/black/brown) can yield ambiguous or noisy spectra \citep{Faltynkova2021}. In hyperspectral imaging, the richness of information comes at a cost: high spectral and spatial resolution can require drastically longer acquisition times, motivating strategies that reduce the number of informative bands without sacrificing identification performance \citep{Yang2024}.

These observations suggest three methodological gaps. First, many workflows rely predominantly on empirical modelling without systematically leveraging \emph{a priori} spectral knowledge (e.g., characteristic absorption regions and known confusable polymer pairs) to guide robust identification under realistic variability \citep{Zhu2023,VallsConesa2023}. Second, uncertainty factors and confusability risks are rarely formalized, even though misclassifications are often linked to identifiable causes such as polymer chemistry similarity, weathering-induced within-class variance, or measurement conditions \citep{Zhu2023,VallsConesa2023}. Third, scalability is often treated implicitly: pixel-wise comparison to reference libraries or computationally heavy modelling across entire images can be prohibitive for routine analysis, especially for large FT-IR and hyperspectral datasets \citep{Faltynkova2021,Yang2024}. While clustering and dimensionality reduction are used in the literature for exploration or preprocessing \citep{Vidal2012,DaSilva2020,Yan2022}, they are less commonly positioned as explicit computational strategies to reduce identification cost while preserving chemically meaningful structure.

To address these gaps, this work proposes a unified multi-level framework that links spectral-level and image-level corrections with scalable modelling and identification strategies. The framework emphasizes (i) robustness to experimental variability, (ii) explicit treatment of confusability and uncertainty (including high-risk pairs such as PP/PE), and (iii) computational scalability through representative modelling and matching at the level of learned spectral structures rather than exhaustive pixel-wise operations.

\subsection{Contributions}

The novel contributions of the present paper are:
\begin{itemize}
  \item We propose a multi-level preprocessing framework that integrates spectral-level and image-level corrections within a unified and scalable workflow for spectral imaging of microplastics.

  \item We introduce a knowledge-informed identification strategy that combines empirical modelling with \emph{a priori} spectral information, explicitly accounting for known confusable polymer classes (e.g., PP and PE).

  \item To ensure scalability, we propose a cluster-based matching approach in which spectral library comparison is performed on representative cluster centroids rather than on individual pixels, substantially reducing computational complexity for large images.

  \item We incorporate an uncertainty-aware analysis of class confusability, identifying experimental factors that increase misclassification risk (e.g. acquisition variability) and evaluating their impact on identification performance.
\end{itemize}

\section{Materials and methods}\label{ssec:matmet}

\subsection{Data acquisition}

The present paper utilised both publicly available FT-IR polymer spectra, along with in-house acquired observations. The public datasets comprised labelled FT-IR spectra acquired under different experimental conditions and instrumentation setups, including both pristine and environmentally altered polymer samples. The datasets contained spectra from the major polymer classes considered in this study: polystyrene (PS), polyethylene terephthalate (PET), polyetylene (PE), and polypropylene (PP), and were used for supervised model calibration, validation, and construction of the spectral reference library. Prior to modelling, all datasets were harmonized to a common spectral range and spectral resolution through interpolation and preprocessing. The combined FT-IR spectral library covered the spectral range from 4000 to 600~cm$^{-1}$ and consisted of 1876 PE spectra, 501 PET spectra, 1015 PP spectra, and 573 PS spectra obtained from three datasets: Villegas \textit{et al.}~\cite{villegas2024ftir}, Kedzierski \textit{et al.}~\cite{Kedzierski2019}, and Jung \textit{et al.}~\cite{jung2018polymer}. They were divided into 70\% training spectra (2776 samples) and 30\% testing spectra (1181 samples), which were harmonised to 301 wavenumbers with 8.0 cm$^{-1}$ spectral resolution. 

For further transferability testing and semi-supervised procedure development, in-situ data was acquired. A sample was prepared using particles of known polymers from four polymer types: PS, PET, PE, and PP. The size of the particles range from 16.5 $\mu$m to 1165.6 $\mu$m. The particles were transferred onto a gold-coated filter membrane  (5 $\mu$m pore size) with the help of pre-cleaned tweezers, under a steromicroscope for subsequent FT-IR analysis. The known-polymer samples consisted of pristine PE, PET, PP, and PS materials obtained directly from the producer. The samples were neither environmentally collected nor artificially aged. Their polymer identities were based on the information supplied by the producer and were independently confirmed using ATR-FTIR before the FT-IR imaging measurements.

FT-IR analysis was performed with the Agilent Cary 670/620 imaging spectrometer, with a 128X128 FPA detector. The measurements were made in reflection mode, and the entire filtered area was analyzed. The measurement settings were as follows: 15X objective, 5.5 $\mu$m pixel size, 3800–800 cm$^{-1}$ spectral range, a spectral resolution of 8 cm$^{-1}$, and 4 number of scans per measurement. Before modelling, the spectra from the three public datasets and the in-house FT-IR image spectra were mapped by linear interpolation onto a common wavenumber grid spanning 800--3200~cm$^{-1}$ at 8~cm$^{-1}$ intervals. Consequently, all spectra supplied to the models contained the same 301 wavenumber variables.

%Analysis was performed with imaging FTIR spectroscopy (Agilent Cary 670/620, 128X128 FPA detector). The measurements were made in reflection mode and the entire filtered area was analyzed. The measurement settings were as follows: 15X objective, 5.5 $\mu$m pixel size, 3800–800$^{-1}$ spectral range, spectral resolution 8 cm$^{-1}$, and 4 number of scans. The results were calculated with the siMPle program. It calculates the number and size of microplastic particles from the infrared spectrum map and identifies different polymer types. The analysis examined the most common types of plastic. The smallest possible recognizable microplastic size is around 10–20 $\mu$m.

\subsection{Overview of the proposed workflow}

\begin{figure}[H]
    \centering
    \includegraphics[width=1\linewidth]{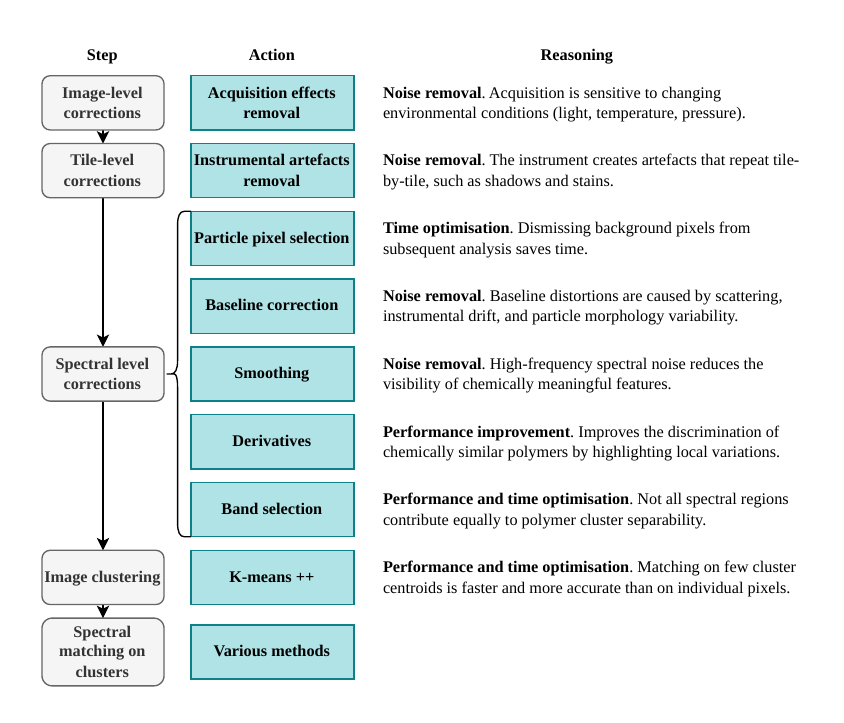}
    \caption{The processing steps considered for the multi-level framework in the case of a semi-supervised particle identification approach, along with the reasoning.}
    \label{fig:workflow}
\end{figure}

The present paper looks at the improvements in particle identification from two perspectives (a) from a semi-supervised approach, and (b) in a supervised approach, using classification approaches. The image-level, tile-level, and spectral-level corrections tested are identical for either the two approaches, and are mathematically presented in the present section. Figure~\ref{fig:workflow} gives an overview on the pre-processing considered, and their reasoning, which is subsequently described in the following sections. 

To reduce the computational burden, after tile-level corrections, spectral preprocessing and subsequent particle-identification operations were applied only to pixels retained after foreground/background separation, rather than to all pixels in the hyperspectral image. The computational workload of these stages therefore depends principally on the number of candidate particle pixels, which is determined by both image size and particle coverage. This particle-pixel-wise strategy avoids unnecessary processing of background spectra and contributes to the comparatively short processing time reported for the analysed image.

\subsubsection{Spectral image representation}

A spectral image is represented as a three-way array \(\mathcal{X} \in \mathbb{R}^{H \times W \times B} \),
where $H$ and $W$ denote the spatial dimensions and $B$ the number of spectral bands. The spectrum at spatial location $(u,v)$ is denoted by \(
\mathbf{x}_{u,v} = \left[ X(u,v,1),\,x(u,v,2),\,\dots,\,X(u,v,B) \right]^\top \in \mathbb{R}^{B} \).
For matrix-based preprocessing and modelling, the cube may be unfolded into \(\mathbf{X} \in \mathbb{R}^{N \times B}\), where \( N = H W\)
and each row of $\mathbf{X}$ corresponds to one pixel spectrum. The notation utilised throughout the paper can be seen in Table~\ref{tab:notation}.

\begin{table}[H]
\centering
\caption{Image-related notations used throughout the paper.}
\begin{tabular}{ll}
\toprule
Symbol & Description \\
\midrule
$\mathcal{X} \in \mathbb{R}^{H \times W \times B}$ & Spectral image cube \\
$X(u,v,d)$ & Intensity at pixel $(u,v)$ and band $d$ \\
$\mathbf{x}_{i} \in \mathbb{R}^{B}$ & Spectrum at pixel $i$ \\
$B$ & Number of spectral bands \\
$H, W$ & Image height and width \\
$N = H \cdot W$ & Total number of pixels \\
$q = 1,\dots,Q$ & Tile index \\
$(u,v)$ & Coordinates within a tile \\
$m$ & Tile size ($m \times m$) \\
$X_q(u,v,d)$ & Pixel in tile $q$ at band $d$ \\
$M_q(u,v)$ & Binary mask (valid pixels) \\
$A(u,v,d)$ & Tile-repeated artefact \\
$S_q(u,v,d)$ & True signal \\
\bottomrule
\end{tabular}
\label{tab:notation}
\end{table}

Because $N = HW$ may be very large, preprocessing is performed in batches or tiles. Let $\mathbf{X}^{(q)} \in \mathbb{R}^{N_q \times B}$ denote batch $q$, with
\begin{equation*}
\mathbf{X} = 
\begin{bmatrix}
\mathbf{X}^{(1)}\\
\mathbf{X}^{(2)}\\
\vdots\\
\mathbf{X}^{(Q)}
\end{bmatrix},
\qquad
\sum_{q=1}^{Q} N_q = N.
\end{equation*}
This representation allows memory-efficient processing without loading the full unfolded image into memory simultaneously.

\subsection{Multi-level preprocessing}

\subsubsection{Spectral level corrections}

Spectral-level corrections aim to mitigate distortions that affect individual pixel spectra prior to spatial processing or modelling. In microplastic analysis using FT-IR and hyperspectral imaging, such distortions commonly arise from baseline drift, scattering effects, limited spectral resolution, additive noise, and variability in particle thickness or surface morphology \citep{DaSilva2020,Faltynkova2021,Nicolau2024}. In environmental samples, additional variability may result from weathering, fouling, and the presence of additives, which can suppress characteristic peaks or introduce spurious features \citep{Yan2022,Zhu2023}. These factors increase within-class variance and contribute to class confusability, paricularly for the polymers that have a similar chemical fingerprint signature, such as the PE and the PP \citep{Zhu2023,VallsConesa2023}.

Baseline distortions are frequently observed in FT-IR spectra due to scattering and instrumental effects. A variety of approaches have been reported in the literature, including asymmetric least squares (AsLS), adaptive iteratively reweighted penalized least squares (airPLS), polynomial detrending, and joint baseline–signal decomposition strategies \citep{Yan2022,DaSilva2020,VallsConesa2023}. In particular, methods such as Fast Background Correction and Identification (FBCI) perform simultaneous estimation of reference signal and baseline components, thereby reducing the risk of over- or under-fitting compared to separate baseline removal procedures \citep{VallsConesa2023}. Robust baseline correction is essential to preserve chemically meaningful absorption bands, especially in the fingerprint region where subtle differences may discriminate confusable polymers.

Let $\mathbf{x}_i \in \mathbb{R}^{B}$ denote the spectrum of pixel $i$. Spectral-level preprocessing begins by modelling the measured spectrum as
\begin{equation}\label{eq:measuredSpectrum}
\mathbf{x}_i = \mathbf{s}_i + \mathbf{b}_i + \boldsymbol{\varepsilon}_i,
\end{equation}
where $\mathbf{s}_i$ is the chemically relevant signal, $\mathbf{b}_i$ is a slowly varying baseline component, and $\boldsymbol{\varepsilon}_i$ is noise. After estimating the baseline $\widehat{\mathbf{b}}_i$, the corrected spectrum is
\begin{equation}\label{eq:baselineCorrectedSpectrum}
\mathbf{x}_i^{\mathrm{bc}} = \mathbf{x}_i - \widehat{\mathbf{b}}_i.
\end{equation}

In the case of AsLS, the baseline estimate is obtained as
\begin{equation}\label{eq:asls}
\widehat{\mathbf{b}}_i
=
\arg\min_{\mathbf{b}}
\left[
\sum_{j=1}^{B} w_{ij}\left(x_{ij}-b_j\right)^2
+
\lambda \sum_{j=2}^{B-1} (\Delta^2 b_j)^2
\right],
\end{equation}
where $\lambda$ controls smoothness, $w_{ij}$ are asymmetry weights, and $\Delta^2$ denotes the second-order finite difference operator.

Additive and multiplicative intensity variations can arise from differences in particle thickness, morphology, and surface roughness. Standard normalization strategies, such as standard normal variate (SNV), min--max scaling, and z-score normalization, are commonly applied to reduce inter-spectrum variability \citep{DaSilva2020,Yan2022}. These corrections are particularly relevant in environmental samples, where large within-class variance due to non-uniform weathering and size effects has been reported \citep{VallsConesa2023}. By stabilizing intensity variations, normalization enhances the comparability of spectra across pixels and acquisition conditions.

The SNV can be expressed as:
\begin{equation}\label{eq:SNV}
\mathbf{x}_i^{\mathrm{snv}} =
\frac{
\mathbf{x}_i^{\mathrm{bc}} - \mu_i 
}{
\sigma_i
},
\end{equation}
where $\mu_i$ and $\sigma_i$ are the mean and standard deviation of the spectrum $\mathbf{x}_i^{\mathrm{bc}}$.

Spectral noise, especially in low-reflectance particles (e.g., dark or weathered MPs), can obscure diagnostic peaks and impair classification \citep{Faltynkova2021}. Smoothing methods, such as Savitzky--Golay (SG) filtering, are commonly employed to suppress high-frequency noise while preserving peak shape \citep{DaSilva2020}. Derivative transformations and peak-search algorithms have also been used to enhance spectral features and improve discrimination in overlapping spectral regions \citep{Yan2022}. However, excessive smoothing may distort subtle spectral features, whereas derivative transformations can amplify high-frequency noise, particularly when the spectral resolution or signal-to-noise ratio is limited \citep{Nicolau2024}. Careful parameter selection is therefore required to balance noise suppression, feature enhancement, and the preservation of discriminative spectral information.

Optional smoothing can be written as
\begin{equation}\label{eq:SGfilt}
\mathbf{x}_i^{\mathrm{sm}} = \mathbf{H}\,\mathbf{x}_i^{\mathrm{snv}},
\end{equation}
where $\mathbf{H}$ is a linear smoothing operator, such as a Savitzky--Golay filter.

Principal component analysis (PCA) has been widely used for noise reduction and background suppression \citep{Vidal2012,DaSilva2020}. More recently, targeted band-selection strategies have been proposed to reduce acquisition and computational costs. For example, key discriminative wavenumbers have been identified based on model feature importance, such as SVM-derived importance scores, to enhance nylon MP detection \citep{Yang2024}. Reducing the number of acquired bands can decrease the acquisition time of hyperspectral systems \citep{Yang2024}.

For noise reduction or reduced spectral representation, the preprocessed data matrix may be approximated using a low-rank PCA model and a residual matrix:
\begin{equation}\label{eq:PCArepr}
\mathbf{X} = \mathbf{T}\mathbf{P}^{\top} + \mathbf{E},
\end{equation}
where $\mathbf{X}\in\mathbb{R}^{N\times B}$ is the preprocessed spectral data matrix containing $N$ spectra and $B$ spectral bands, $\mathbf{T}\in\mathbb{R}^{N\times K_{\mathrm{keep}}}$ is the score matrix, $\mathbf{P}\in\mathbb{R}^{B\times K_{\mathrm{keep}}}$ is the corresponding loading matrix, $K_{\mathrm{keep}}$ is the number of retained principal components, and $\mathbf{E}\in\mathbb{R}^{N\times B}$ is the residual matrix.

In the proposed framework, spectral-level corrections are designed not only to improve overall classification performance but also to explicitly reduce within-class variance and enhance separability for known confusable polymer pairs. By integrating baseline correction, normalization, smoothing, and informed band selection in a structured manner, the spectral-level stage provides stabilized and information-rich inputs for subsequent clustering, modelling, and uncertainty-aware identification.

\subsubsection{Image level corrections}

While spectral-level corrections address distortions within individual spectra, image-level corrections target spatially structured variability arising during acquisition. In FT-IR and hyperspectral imaging, images are often acquired tile-by-tile over extended periods, making them susceptible to temporal drift in environmental conditions (e.g., illumination fluctuations, temperature changes, background reflectance variability) \citep{Faltynkova2021}. Such variability can introduce systematic shifts across tiles, leading to spatial inconsistencies that degrade classification robustness and inflate within-class variance.

Let $\mathbf{X}_{\mathrm{bg}} \in \mathbb{R}^{N_{\mathrm{bg}} \times B}$ denote spectra from background regions. PCA yields the decomposition
\begin{equation}\label{eq:PCAbg}
\mathbf{X}_{\mathrm{bg}} = \mathbf{T}_{\mathrm{bg}}\mathbf{P}^\top_{\mathrm{bg}} + \mathbf{E}_{\mathrm{bg}},
\end{equation}
where $\mathbf{T}_{\mathrm{bg}}$ contains principal component scores, $\mathbf{P}_{\mathrm{bg}}$ loading vectors, and $\mathbf{E}_{\mathrm{bg}}$ residuals.

Four fixed regions of interest (ROIs) were selected from particle-free areas of the background plate and retained throughout the analysis. As these ROIs represented unchanged background material, their PC-score distributions were expected to remain comparatively stable during acquisition under consistent measurement conditions.

For principal component $\ell$ and background region $r$, let
\begin{equation}
\mathcal{T}_{\ell}^{(r)}
=
\left\{t_{i\ell}:i\in\mathcal{I}_r\right\}
\end{equation}
denote the set of scores for PC $\ell$ within background region $r$, where $\mathcal{I}_r$ is the set of pixel indices belonging to ROI $r$.

For each principal component, the empirical score distributions in the fixed background ROIs were compared across the image tiles using the Wasserstein distance \cite{panaretos2019statistical}. The resulting distances were evaluated in tile-acquisition order. A PC was considered a nuisance-component candidate when its background-score distributions exhibited a pronounced and systematic temporal pattern across the tiles, rather than isolated differences among individual pixels or ROIs. No formal statistical significance test or $p$-value threshold was applied. Instead, the Wasserstein distance was used as a quantitative measure of distributional dissimilarity, and nuisance-PC identification was performed diagnostically based on both the magnitude and temporal structure of the observed changes.

Following identification of the nuisance components, the spectral image was reconstructed using the retained principal components:
\begin{equation}
\widehat{\mathbf{X}}
=
\sum_{\ell\in\mathcal{K}_{\mathrm{keep}}}
\mathbf{t}_{\ell}\mathbf{p}_{\ell}^{\top},
\end{equation}
or, equivalently,
\begin{equation}
\widehat{\mathbf{X}}
=
\mathbf{T}_{\mathrm{keep}}
\mathbf{P}_{\mathrm{keep}}^{\top},
\end{equation}
where $\mathcal{K}_{\mathrm{keep}}$ is the set of retained principal components, $\mathbf{t}_{\ell}$ and $\mathbf{p}_{\ell}$ are the score and loading vectors for PC $\ell$, respectively, and $\mathbf{T}_{\mathrm{keep}}$ and $\mathbf{P}_{\mathrm{keep}}$ contain the scores and loadings of the retained components.

PCA-based reconstruction provides a low-rank representation in which the identified nuisance components are suppressed while the dominant chemically relevant spectral structure is retained \citep{Vidal2012,DaSilva2020}. In the present workflow, this reconstruction is used to reduce temporally structured background variation across tiles and thereby limit artificial within-class variability.

In addition to variance-based ordering, noise-oriented transforms such as the Minimum Noise Fraction (MNF) transform have been proposed to rank components according to signal-to-noise characteristics \citep{Bhargava2000}. Although MNF is commonly used for morphology visualization and noise suppression, the present framework leverages PCA in combination with background monitoring to explicitly disentangle environmental drift from chemical variability.

Spectral imaging data may also contain structured artefacts, including dead pixels, striping patterns, and spectral spikes caused by sensor or detector irregularities \citep{Faltynkova2021}. Such artefacts can produce spatially coherent false positives or distort particle boundaries during segmentation. Artefact mitigation is therefore performed prior to modelling using a combination of outlier detection, local smoothing, and spatial filtering. 

Where necessary, morphological filtering and connected-component constraints are applied to prevent isolated noisy pixels from being classified as particles, consistent with recommendations in the literature to reduce false positives and misclassification of adjacent particles \citep{Faltynkova2021}. By integrating artefact reduction with condition-aware reprojection, the proposed image-level correction stage enhances both spatial coherence and chemical interpretability prior to clustering and identification.

\subsubsection{Tile-level corrections}

In addition to global image-level variability, spectral images acquired in a tile-by-tile manner may exhibit tile-repeated fixed-pattern artefacts arising from detector non-uniformities and stitching effects. These artefacts are assumed to be stationary in tile coordinates and repeat across all tiles, while the underlying sample signal varies spatially. This property enables estimation of the artefact component by aggregating information across tiles.

Let $X_q(u,v,d)$ denote the observed intensity at within-tile coordinates $(u,v)$, spectral band $d$, and tile index $q=1,\dots,Q$. The observation is modelled as
\begin{equation}\label{eq:pixelInTile}
X_q(u,v,d) = S_q(u,v,d) + A(u,v,d) + \varepsilon_q(u,v,d),
\end{equation}
where $S_q(u,v,d)$ represents the true sample signal, $A(u,v,d)$ is a tile-repeated artefact fixed in tile coordinates, and $\varepsilon_q(u,v,d)$ denotes noise.

To reduce the influence of tile-wise intensity offsets, each tile is detrended prior to artefact estimation. Using median detrending, the residual is defined as
\begin{equation} 
R_q(u,v,d) = X_q(u,v,d) - \mu_q^{(d)},
\end{equation}
where \( \mu_q^{(d)} = \operatorname{median}\{X_q(u,v,d) : M_q(u,v)=1\} \)
and $M_q(u,v)$ is a binary mask defining valid pixels (e.g., background).

The tile artefact template is estimated by averaging residuals across tiles:
\begin{equation}
\widehat{A}(u,v,d)
=
\frac{
\sum_{q=1}^{Q} M_q(u,v)\,R_q(u,v,d)
}{
\sum_{q=1}^{Q} M_q(u,v)
}.
\end{equation}

To ensure that only spatially structured artefacts are removed, the template is centered to zero mean for each band:
\begin{equation}
\widehat{A}_0(u,v,d) = \widehat{A}(u,v,d) - \overline{\widehat{A}}^{(d)},
\end{equation}
where $\overline{\widehat{A}}^{(d)}$ is the band-wise mean.

The corrected image is obtained by subtracting the estimated artefact template from each tile:
\begin{equation}\label{eq:tileCorrection} 
X_q^{\mathrm{corr}}(u,v,d) = X_q(u,v,d) - \widehat{A}_0(u,v,d).
\end{equation}

By restricting the estimation to background pixels, the influence of spatially localized sample structures is reduced, ensuring that the estimated template captures detector- and acquisition-related artefacts rather than true chemical variation.

\subsection{Modelling strategies}

\subsubsection{Unsupervised modelling with library search}

Unsupervised modelling is employed to explore latent structure in spectral images without requiring exhaustive pixel-level labels. Clustering and latent variable techniques such as PCA, non-negative matrix factorization (NMF), and matrix factorization approaches have been used in hyperspectral MP studies to reveal material groupings and suppress background variability \citep{Vidal2012,DaSilva2020,Yan2022}. 

In the proposed framework, clustering serves a dual purpose. First, it identifies spectrally homogeneous regions that likely correspond to individual materials or particle segments. Second, it provides a computationally efficient abstraction of the image: instead of performing spectral library comparison or complex modelling on millions of individual pixels, representative cluster centroids are extracted and used for subsequent matching and classification. This centroid-based strategy reduces computational burden while preserving chemically meaningful structure, thereby supporting scalability for large FT-IR image datasets.

Unsupervised analysis also facilitates detection of outliers and potentially novel or highly weathered materials that deviate from known polymer classes. Clusters exhibiting large residuals with respect to reference models or low similarity to library spectra can be flagged for further inspection.

Let $\{\mathcal{C}_1,\dots,\mathcal{C}_K\}$ denote a partition of the spectra into $K$ clusters. The centroid of cluster $k$ is defined as
\begin{equation}\label{eq:clusterCentroids}
\boldsymbol{\mu}_k = \frac{1}{|\mathcal{C}_k|} \sum_{i \in \mathcal{C}_k} \mathbf{x}_i.
\end{equation}

Instead of comparing all $N$ spectra to a reference library, matching may be performed on the set of centroids $\{\boldsymbol{\mu}_k\}_{k=1}^{K}$ with $K \ll N$, thereby reducing computational cost.

\begin{table}[H]
\centering
\scriptsize
\setlength{\tabcolsep}{3pt} % reduce column spacing
\renewcommand{\arraystretch}{1.1} % slightly tighter rows

\begin{tabularx}{\linewidth}{p{0.04\linewidth} p{0.14\linewidth} X p{0.08\linewidth}}
\hline
\textbf{M.} & \textbf{Name} & \textbf{Description} & \textbf{Ref.} \\
\hline

M1 & Peak positions &
Compares spectra using the locations of prominent bands, emphasizing diagnostic absorptions while being less sensitive to intensity scaling. &
\cite{brown2014curvefit} \\

M2 & Peak ratios &
Uses ratios of selected band intensities at diagnostic wavenumbers, capturing relative peak relationships instead of absolute amplitudes. &
\cite{corrado2025lithium} \\

M3 & Band integrals &
Integrates signal over predefined spectral windows, summarizing broad absorption behaviour rather than individual peaks. &
\cite{shanmugam2014spectral}  \\

M4 & Multi-region cosine similarity &
Computes cosine similarity across multiple spectral regions and averages the scores to balance agreement across chemically meaningful windows. &
\cite{vandermeer2006spectral} \\

M5 & Sign-invariant derivative cosine &
Compares first-derivative spectra using cosine similarity and selects the better of normal and sign-flipped comparisons, making it robust to derivative sign mismatches. &
\cite{savitzky1964smoothing,vandermeer2006spectral, galal2012novel} \\

M6 & DTW on derivatives &
Applies dynamic time warping to first-derivative spectra, allowing local spectral shifts before computing dissimilarity. &
\cite{sakoe1978dynamic} \\

M7 & SAM &
Measures the angle between spectra as vectors, focusing on spectral shape and being insensitive to scale. &
\cite{kruse1993sips} \\

M8 & Weighted cosine &
A cosine similarity where selected spectral regions receive higher weights to emphasize diagnostically important bands. &
\cite{he2011wsam} \\

M9 & PCA matching &
Projects spectra into principal-component space and compares them in a reduced-dimensional representation. &
\cite{pearson1901lines,jolliffe2002pca,vane1988pca} \\

M10 & KNN vote &
Assigns class labels based on nearest neighbours in the reference library using a voting scheme. &
\cite{cover1967nearest, melgani2004classification} \\

M11 & Raw + derivative correlation &
Combines correlation scores from raw and derivative spectra to exploit both global and local spectral agreement. &
\cite{pearson1909correlation,savitzky1964smoothing, tsuchikawa2007review} \\

M12 & Continuum cosine &
Removes the spectral continuum and compares normalized spectra using cosine similarity to emphasize feature shape. &
\cite{clark1984reflectance, ren2020improved} \\

\hline
\end{tabularx}

\caption{Summary of the twelve spectral-library matching methods evaluated in this work.}
\label{tab:matching_methods}
\end{table}

\subsubsection{Supervised modelling}

Supervised modelling aims to assign each pixel or region to a predefined polymer class using labelled training data. In the microplastics literature, widely adopted methods include partial least squares discriminant analysis (PLS-DA), support vector machines (SVM), k-nearest neighbours (KNN), random forests (RF), soft independent modelling of class analogy (SIMCA), and ensemble approaches \citep{Kedzierski2019,Hufnagl2019,Michel2020,DaSilva2020,Yan2022,VallsConesa2023}. These methods have demonstrated improved average classification performance compared to conventional library search, particularly under controlled acquisition conditions.

However, many commonly used approaches are linear in nature (e.g., classical PLS-DA), and may therefore struggle when spectral differences between polymers are subtle or nonlinear, as is the case for chemically similar classes such as PE and PP. While PLS-DA has been shown to outperform some spectral matching strategies \citep{Faltynkova2021}, extensions that incorporate nonlinear structure—such as kernel-based PLS-DA—remain largely unexplored in microplastic spectral imaging.

In this framework, supervised modelling is formulated as a comparative stage that includes both conventional linear multivariate classifiers and nonlinear extensions. The objective is not only to improve overall accuracy, but to assess whether nonlinear decision boundaries enhance separability for confusable polymer pairs and under realistic environmental variability. Model performance is evaluated using cross-validation and held-out image regions to reduce spatial leakage between training and test samples.

Let $\mathbf{x}_i \in \mathbb{R}^{B}$ denote a preprocessed spectrum and let $c_i \in \{1,\dots,C\}$ denote its class label. Supervised modelling seeks a mapping
\begin{equation}
f : \mathbb{R}^{B} \rightarrow \{1,\dots,C\},
\end{equation}
such that
\begin{equation}
\hat{c}_i = f(\mathbf{x}_i),
\end{equation}
where $\hat{c}_i$ is the predicted class label.

For linear discriminative models, class assignment may be based on
\begin{equation}
g_c(\mathbf{x}_i) = \mathbf{w}_c^\top \mathbf{x}_i + b_c,
\end{equation}
with predicted class
\begin{equation}
\hat{c}_i = \arg\max_{c} g_c(\mathbf{x}_i).
\end{equation}

In kernel-based models, the representation is replaced by a similarity mapping
\begin{equation}
K_{ij} = \kappa(\mathbf{x}_i,\mathbf{x}_j),
\end{equation}
where $\kappa(\cdot,\cdot)$ is a kernel function, enabling nonlinear decision boundaries in the original spectral space. A summary of the utilised methods are presented in Table~\ref{tab:supervisedMethods}.

\begin{table}[H]
\centering
\scriptsize
\setlength{\tabcolsep}{4pt}
\renewcommand{\arraystretch}{1.15}

\caption{Supervised classification methods evaluated for microplastic spectral-image classification.}
\label{tab:supervisedMethods}

\begin{tabularx}{\linewidth}{p{0.14\linewidth} p{0.38\linewidth} p{0.34\linewidth} p{0.04\linewidth}}
\hline
\textbf{Method} & \textbf{Description} & \textbf{Optimised Hyperparameters} & \textbf{Ref.} \\
\hline

PLS-DA &
Linear latent-variable classifier maximising covariance between spectra and class labels. Widely used in chemometric spectral classification. &
Number of latent variables &
\cite{wold2001pls,kedzierski2019microplastics} \\

SIMCA &
Class-modelling approach based on independent PCA models for each polymer class, where classification is performed according to the similarity of spectra to each class subspace. &
Number of principal components per class, class threshold &
\cite{wold1976pattern} \\

Linear SVM &
Maximum-margin linear classifier separating classes using a linear hyperplane. &
Box constraint &
\cite{cortes1995support,yan2022microplastics} \\

Gaussian SVM &
Nonlinear support vector machine using a radial basis function (RBF) kernel. &
Kernel scale, box constraint &
\cite{cortes1995support,hufnagl2019microplastics} \\

KNN &
Assigns class labels according to the nearest training spectra in feature space. &
Number of neighbours, distance metric &
\cite{cover1967nearest,michel2020microplastics} \\

Decision Tree &
Recursive partitioning classifier based on hierarchical threshold rules. &
Maximum splits, split criterion &
\cite{breiman1984classification} \\

Random Forest &
Ensemble of decision trees trained on bootstrapped samples with random feature selection. &
Number of learners, maximum splits &
\cite{breiman2001random} \\

Boosted Trees &
Sequential ensemble of weak decision-tree learners emphasizing previously misclassified samples. &
Number of learners, learning rate &
\cite{friedman2001greedy} \\

Linear Discriminant Analysis (LDA) &
Projects spectra to maximise between-class variance relative to within-class variance. &
Discriminant type &
\cite{fisher1936use} \\

Quadratic Discriminant Analysis (QDA) &
Extension of LDA allowing class-specific covariance structures and quadratic decision boundaries. &
Regularization parameter &
\cite{hastie2009elements} \\

Naive Bayes &
Probabilistic classifier assuming conditional independence between spectral variables. &
Distribution type, kernel smoothing &
\cite{rish2001empirical} \\

Neural Network &
Multilayer nonlinear classifier trained using iterative optimization of network weights. &
Hidden-layer size, regularization &
\cite{goodfellow2016deep} \\

\hline
\end{tabularx}
\end{table}

\section{Results and discussion}

\subsection{Image-level corrections}

At the image level, PCA serves both as a noise-reduction tool and as a mechanism for identifying structured variability related to acquisition artefacts. Low-information components can be discarded to suppress random noise, while components capturing non-chemical variation can be explicitly identified and removed (Eq.~\eqref{eq:PCArepr}).

\begin{figure}[H]
    \centering
    \begin{subfigure}[b]{0.32\linewidth}
        \centering
        \includegraphics[width=0.8\linewidth]{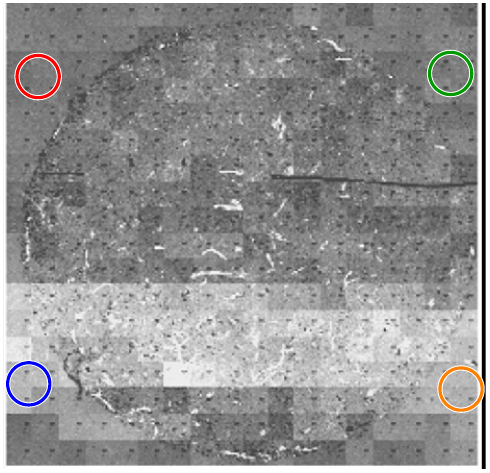}
        \caption{PC\#7 scores ($\mathbf{t}_7$), 1.9\% explained variance}
        \label{fig:imageWithCircles}
    \end{subfigure}
    \begin{subfigure}[b]{0.32\linewidth}
        \centering
        \includegraphics[width=\linewidth]{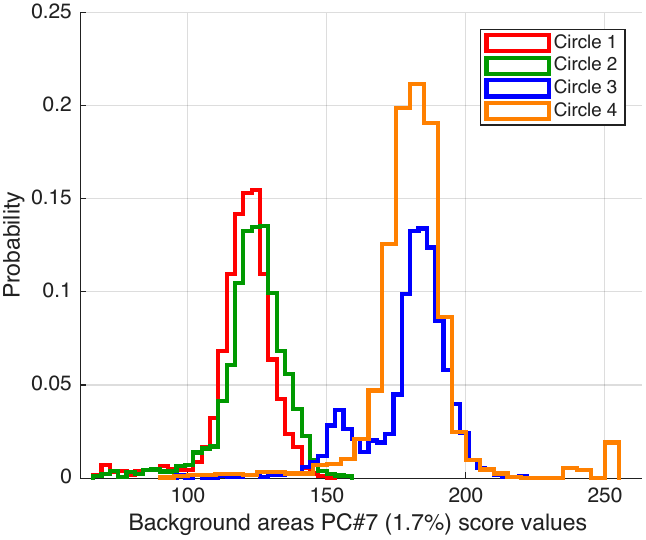}
        \caption{Circled score distributions}
        \label{fig:scoreDistributions}
    \end{subfigure}
    \begin{subfigure}[b]{0.32\linewidth}
        \centering
        \includegraphics[width=\linewidth]{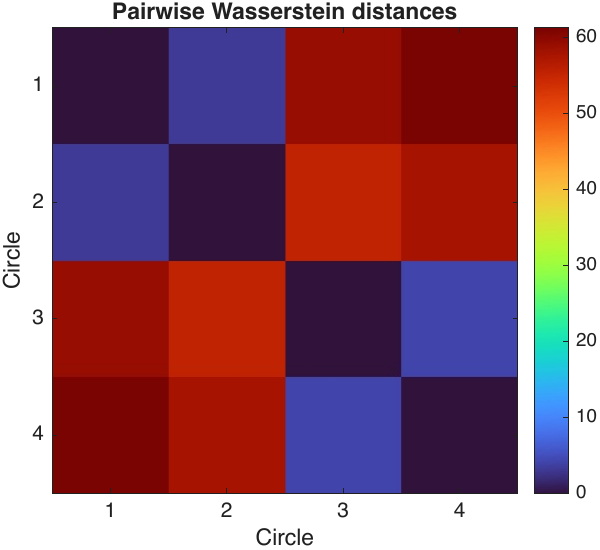}
        \caption{Wasserstein indicator}
    \end{subfigure}
    \caption{The change in environmental conditions is reflected in PC\#7 of this sample, that explains the variance between the initial environmental conditions and the latter ones. Four areas in the PC score that represent non-particle background have been flagged (a). Their score distributions (b) shows a change between the first timeframe of the acquisition, compared to the latter one. The distribution similarity indicator (c) Pairwise Wasserstein distances between the ROI score distributions quantify the magnitude of these shifts. PC\#7 exceeded the empirical distance threshold of 5 and was therefore identified as a nuisance component for removal.}
    \label{fig:Figure1imageDistorsions}
\end{figure}

In the present dataset, an example of such structured variability is shown in Fig. \ref{fig:Figure1imageDistorsions}. Although not all images exhibited this behaviour, this case illustrates the impact of changing environmental conditions during image acquisition. The seventh principal component (PC\#7), explaining 1.9\% of the total variance, was found to capture temporal variability unrelated to sample chemistry.

As shown in Fig. \ref{fig:imageWithCircles}, PC\#7 scores display systematic spatial patterns across the image. Four regions corresponding to non-particle background were selected and analysed. The corresponding score distributions (Fig. \ref{fig:scoreDistributions}) exhibit a clear temporal drift, with values increasing during the intermediate acquisition period and returning towards their initial levels at the end. This temporally structured variation affected the measured sample spectra and was therefore interpreted as an unwanted background or acquisition-related disturbance. Its precise physical origin cannot be determined from the present data: possible explanations include variation in environmental or instrumental conditions, sample heterogeneity evolving across the measurement sequence, or an interaction between these factors. Accordingly, the observed drift should not be regarded as uniquely demonstrating an environmental effect. The correction applied here is based on the temporal structure of the unwanted variation rather than on a definitive causal attribution.

The distributional differences were quantified using pairwise Wasserstein \cite{panaretos2019statistical} distances (Fig. \ref{fig:Figure1imageDistorsions}c), which showed large deviations across all selected background regions. In contrast, no other principal components exceeded the empirical threshold of 5 for this metric. Consequently, PC\#7 was identified as a nuisance component and excluded from subsequent reconstruction and analysis. Removing such components reduces artificial variability and improves the stability of downstream spectral matching and classification. An extended analysis is presented in \textbf{\ref{app:tdi}}.

\subsection{Tile-level corrections}

\begin{figure}[H]
    \centering

    \begin{subfigure}[b]{0.32\linewidth}
        \includegraphics[width=0.82\linewidth]{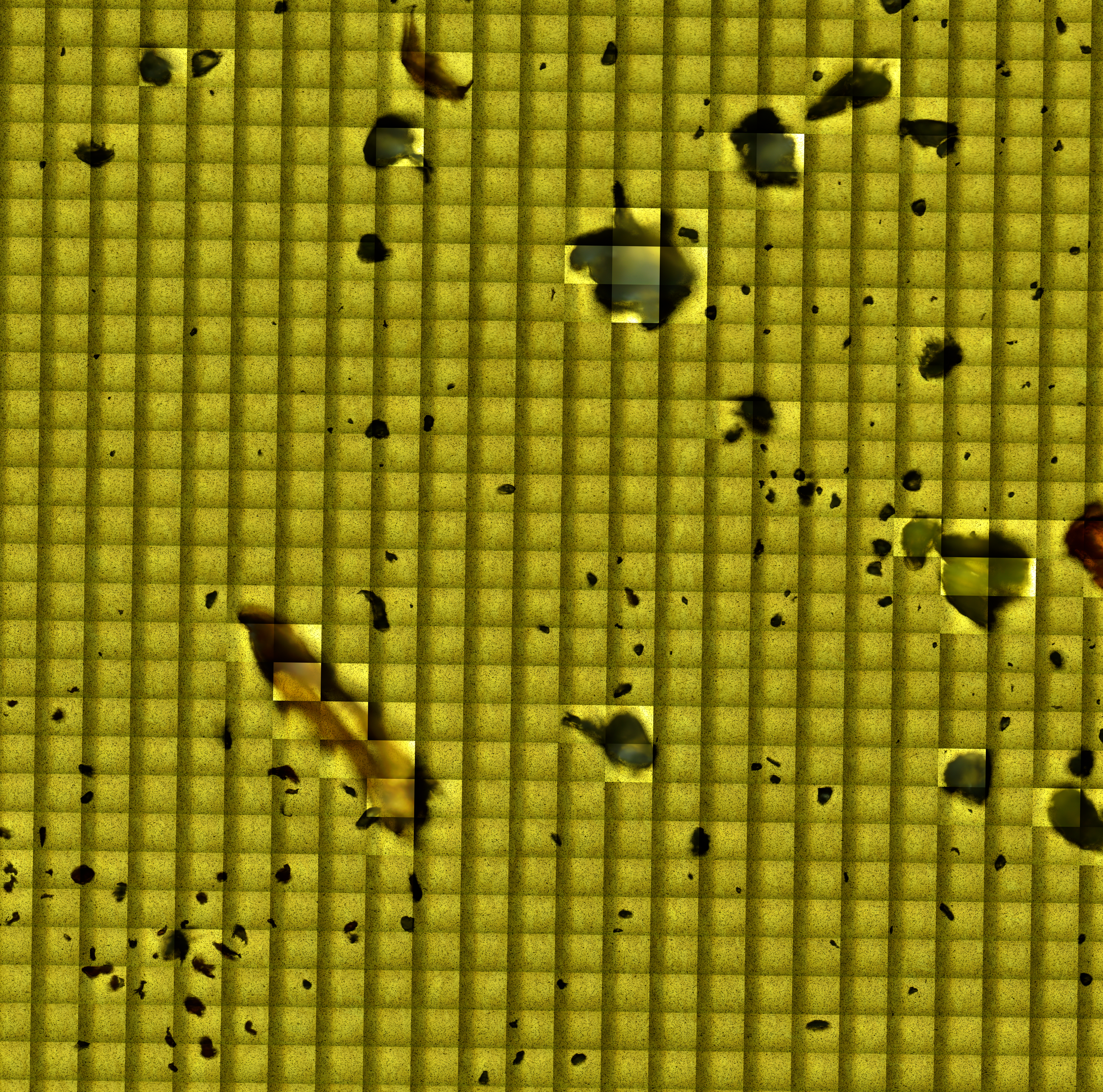}
        \caption{}
        \label{fig:originalVisualisation}
    \end{subfigure}
    \begin{subfigure}[b]{0.32\linewidth}
        \includegraphics[width=\linewidth]{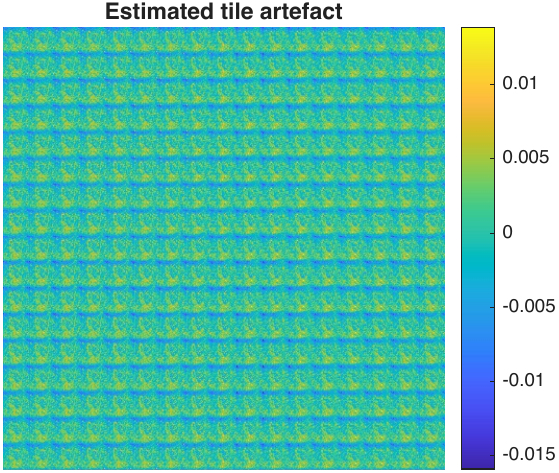}
        \caption{}
        \label{fig:estimatedEffects}
    \end{subfigure}
    \begin{subfigure}[b]{0.32\linewidth}
        \includegraphics[width=0.95\linewidth]{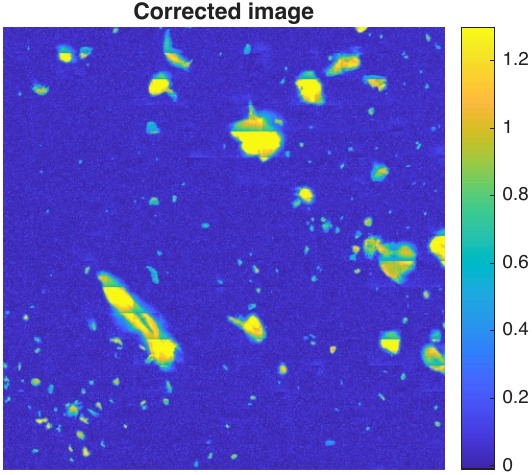}
        \caption{}
        \label{fig:correctedImage}
    \end{subfigure}
    \caption{The original image presenting tile artefacts (a) along with the estimated tile artefacts evaluated with background-only pixels (b). The corrected image is shown in (c) as a visualisation of band averages, and corrects most of these abnormalities, with some effects seen in the largest particles.}
\label{fig:individualTileEffects}
\end{figure}

Tile-dependent artefacts are clearly visible in the raw image (Fig. \ref{fig:individualTileEffects}a), where intensity patterns repeat at the tile scale and manifest as illumination inconsistencies across the field of view. To isolate these effects from chemically meaningful variation, artefact estimation was performed using only background pixels, thereby excluding contributions from microplastic particles, using Eq.~\eqref{eq:tileCorrection}. The resulting artefact template (Fig. \ref{fig:individualTileEffects}b) captures the spatially structured, tile-repeated component associated with acquisition.

After correction, the reconstructed image (Fig. \ref{fig:individualTileEffects}c) exhibits a substantial reduction in tile-wise intensity variation, indicating that the proposed approach effectively mitigates the dominant artefact pattern. However, residual tile-related effects remain observable, particularly within larger particles, where slight intensity discontinuities persist across tile boundaries. This behaviour suggests that part of the variability is not purely additive but may be influenced by tile-wise acquisition conditions, such as local illumination adjustment or interaction with sample morphology.

These remaining discrepancies indicate that tile-level correction alone is insufficient to fully eliminate acquisition-induced variability. Instead, they motivate the complementary use of spectral-level preprocessing, where normalization and derivative-based transformations can further reduce residual intensity inconsistencies and improve robustness in subsequent modelling stages.

\begin{figure}[H]
    \centering
    \begin{subfigure}[b]{0.45\linewidth}
        \centering
        \includegraphics[width=\linewidth]{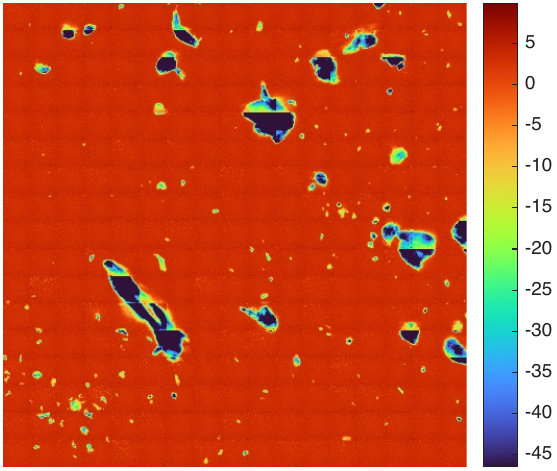}
        \caption{PC \#1 scores ($\mathbf{t}_1$) before tile correction}
        \label{fig:originalPC1}
    \end{subfigure}
    \begin{subfigure}[b]{0.45\linewidth}
        \centering
        \includegraphics[width=0.85\linewidth]{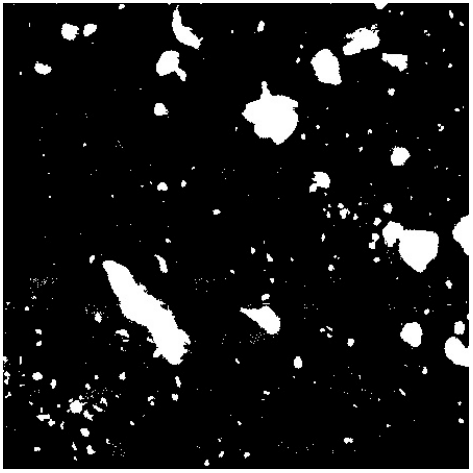}
        \caption{7.55\% of the image pixels where $\mathbf{t}_1 < 0.00$}
        \label{fig:originalMask}
    \end{subfigure}
    \begin{subfigure}[b]{0.45\linewidth}
        \centering
        \includegraphics[width=\linewidth]{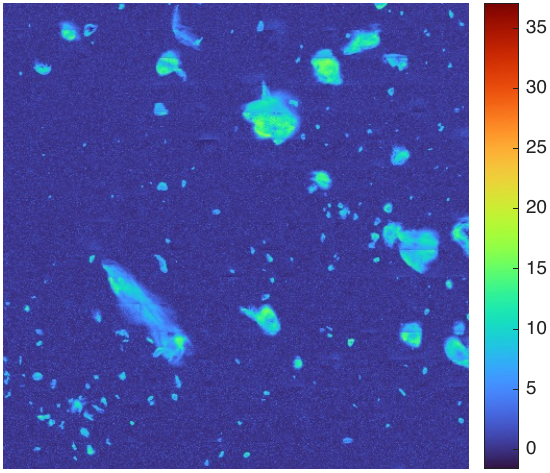}
        \caption{PC\#1 scores ($\textbf{t}_1$) after tile correction}
        \label{fig:correctedPC1}
    \end{subfigure}
    \begin{subfigure}[b]{0.45\linewidth}
        \centering
        \includegraphics[width=0.85\linewidth]{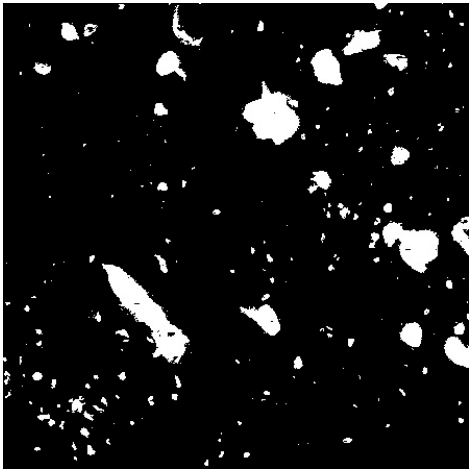}
        \caption{7.00\% of the image pixels  $\mathbf{t}_1 > 0.00$}
        \label{fig:correctedMask}
    \end{subfigure}
    \caption{The 1st PC score, that explains the variance discriminating the particles from the background (a) before and (c) after the tile-by-tile instrumental artefact correction, and their subsequent particle masks (b,d).}
    \label{fig:PCandPixelMasks}
\end{figure}

The impact of tile-level correction is also evident in the particle masking results derived from the first principal component (PC\#1), which consistently captures the dominant variance separating particles from the background (Fig. \ref{fig:PCandPixelMasks}). Prior to correction (Fig. \ref{fig:PCandPixelMasks}a–b), thresholding of PC\#1 scores leads to the appearance of numerous isolated, single-pixel detections, particularly along tile boundaries. These artefacts arise from tile-induced intensity inconsistencies and do not correspond to physically meaningful particles.

Following tile-level correction (Fig. \ref{fig:PCandPixelMasks}c–d), such detections are largely eliminated, resulting in a more spatially coherent and physically realistic particle mask. This demonstrates that removing tile-wise artefacts improves not only visual consistency but also the reliability of downstream segmentation.

Thresholding is performed around the central value of the PC\#1 score. Due to the inherent sign indeterminacy of PCA, the sign of the particle-related scores in PC\#1 may vary between datasets. However, this ambiguity can be resolved by exploiting the class imbalance between background and particle pixels. Since particle pixels typically constitute a small fraction of the image (approximately 7\% in this case), the minority class can be identified by selecting either positive or negative score regions, depending on which corresponds to the smaller proportion of pixels.

After this masking step, only particle spectra are retained for further analysis. Background spectra are excluded from both clustering and spectral matching, as they do not contribute to polymer identification. This reduction in data dimensionality substantially decreases computational cost and improves overall processing efficiency without compromising analytical performance.

\subsection{Spectral-level processing}

\begin{figure}[H]
    \centering
    \begin{subfigure}[b]{0.89\linewidth}
        \centering
        \includegraphics[width=\linewidth]{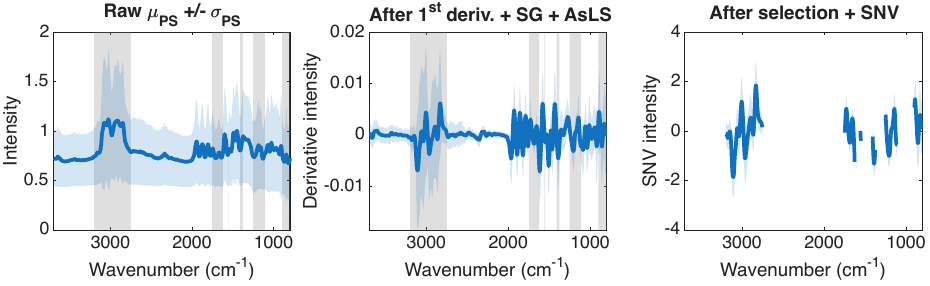}
        \caption{PS}
        \label{fig:PSProcessingEffects}
    \end{subfigure}
    \begin{subfigure}[b]{0.89\linewidth}
        \centering
        \includegraphics[width=\linewidth]{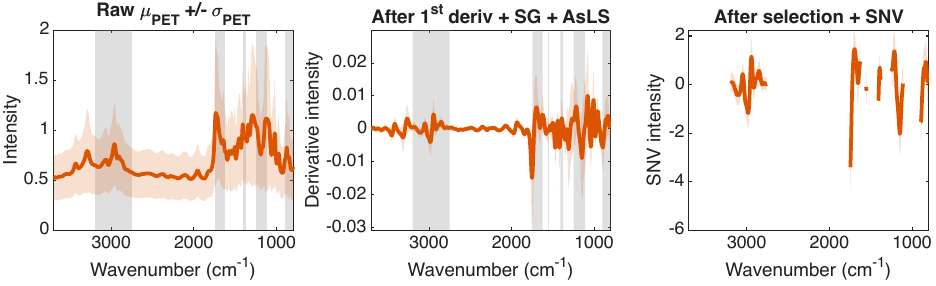}
        \caption{PET}
        \label{fig:PPProcessingEffects}
    \end{subfigure}
    \begin{subfigure}[b]{0.87\linewidth}
        \centering
        \includegraphics[width=\linewidth]{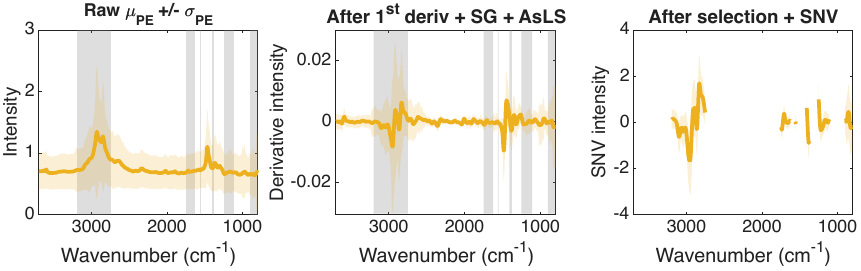}
        \caption{PE}
        \label{fig:PEProcessingEffects}
    \end{subfigure}
    \begin{subfigure}[b]{0.87\linewidth}
        \centering
        \includegraphics[width=\linewidth]{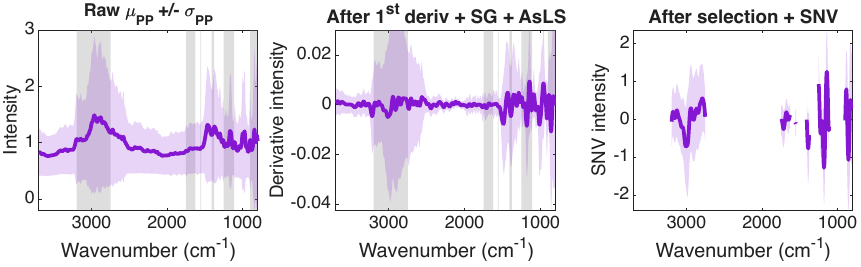}
        \caption{PP}
        \label{fig:PPProcessingEffects}
    \end{subfigure}
    \caption{The raw spectra and the spectral products utilised in the unsupervised modelling phase, illustrated with the (a) PS, (b) PET, (c) PE and (d) PP spectra. The spectral products are: after baseline correction (AsLS, Eq.~\eqref{eq:asls}), smoothing (SG, Eq.~\eqref{eq:SGfilt}) and the first derivative in the intermediary (middle) column, and after wavelengths selection and normalization (SNV, Eq.~\eqref{eq:SNV}) on the right-hand side. }
    \label{fig:preprocessing}
\end{figure}

Spectral preprocessing and variable selection play a critical role in the clustering stage. While clustering is performed on a reduced set of spectral bands, certain preprocessing (Eq.~\eqref{eq:baselineCorrectedSpectrum}, \eqref{eq:asls}, \eqref{eq:SGfilt}) operations—such as smoothing filters and derivative transformations—require the full spectral range (Fig. \ref{fig:PSProcessingEffects}) to ensure consistent and physically meaningful results. For this reason, preprocessing is first applied to the full spectra, after which band selection is performed.

The spectral ranges used for clustering are selected based on literature-reported diagnostic regions for polymer identification \cite{Fakayode2024}. However, it is important to distinguish between wavenumbers that are optimal for clustering and those that are optimal for spectral matching. Clustering benefits from regions that enhance separability between groups in a low-dimensional representation, whereas spectral matching often relies on preserving detailed, chemically specific features across broader or different spectral intervals. As a result, the optimal band selection may differ between these two stages of the workflow.

The effect of spectral preprocessing is illustrated in Fig. \ref{fig:preprocessing}, which presents the mean spectra and corresponding variability (standard deviation) at different stages of processing for each polymer class. The progression from raw spectra to intermediate and final processed representations demonstrates the reduction of baseline effects, normalization of intensity variations, and enhancement of characteristic spectral features. These transformations improve the consistency of intra-class spectra while preserving diagnostically relevant structures, thereby supporting more robust clustering and subsequent identification.

\subsection{Modelling performance}

\subsubsection{Semi-supervised approach}

\begin{figure}[H]
    \centering
    \begin{subfigure}[b]{0.57\linewidth}
        \centering
        \includegraphics[width=0.7\linewidth]{Figure1a.png}
        \caption{}
        \label{fig:theMap}
    \end{subfigure}
    \begin{subfigure}[b]{0.47\linewidth}
        \centering
        \includegraphics[width=0.95\linewidth]{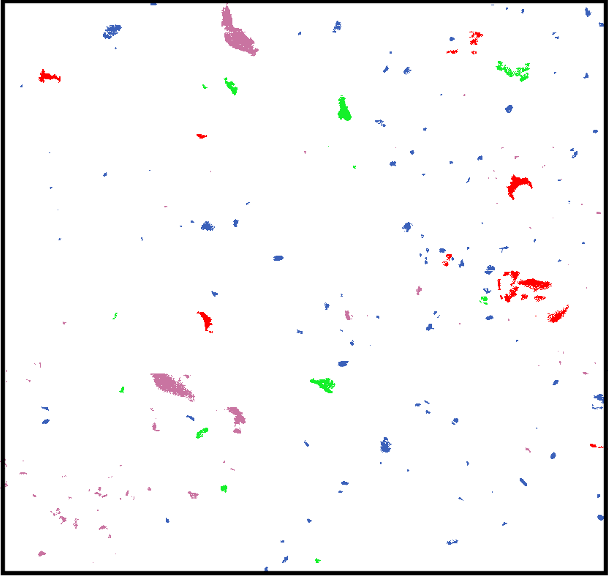}
        \caption{}
        \label{fig:siMPlemap}
    \end{subfigure}
    \begin{subfigure}[b]{0.49\linewidth}
        \centering
        \includegraphics[width=\linewidth]{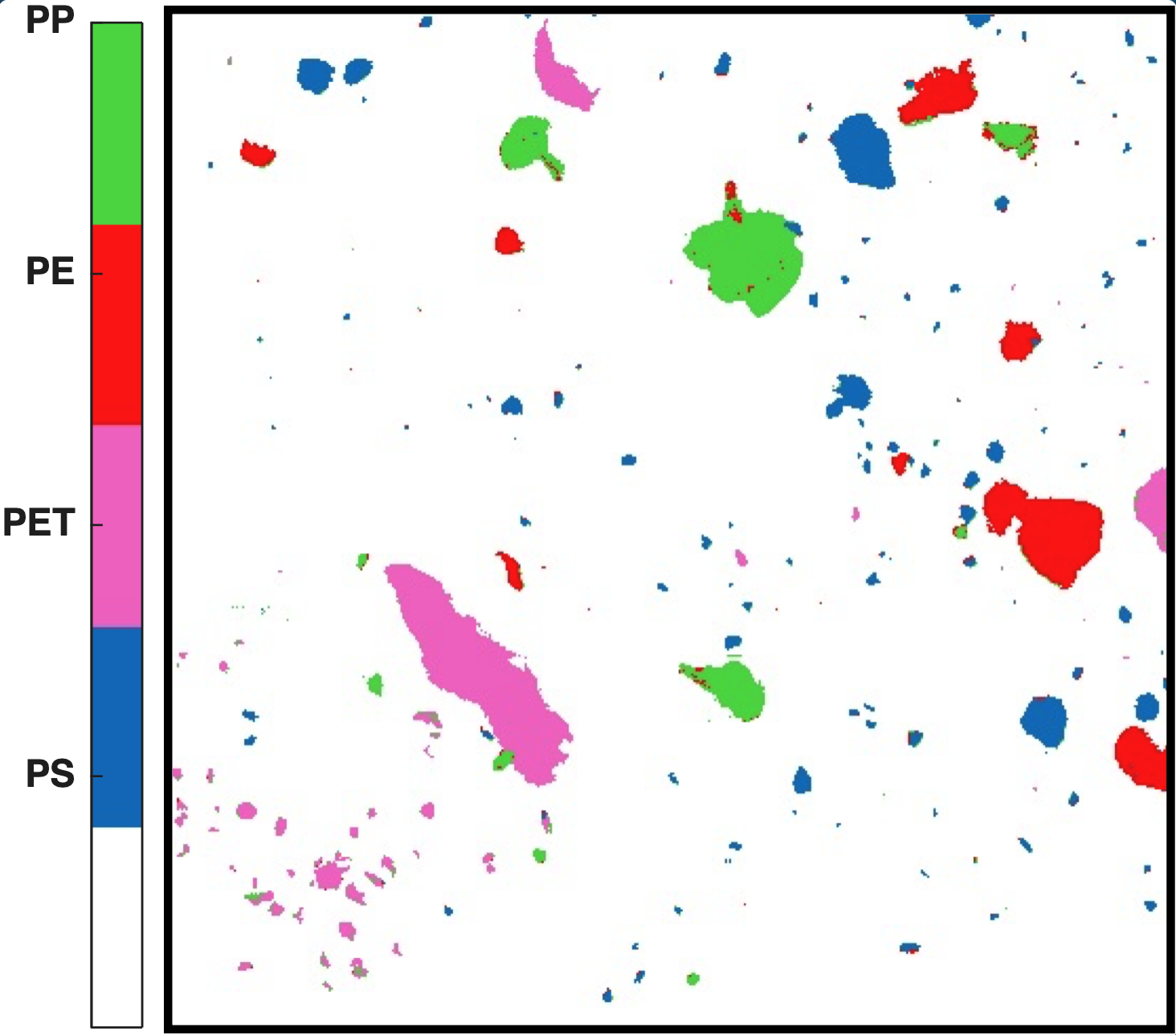}
        \caption{}
        \label{fig:OurMap}
    \end{subfigure}
    \caption{Comparison between (a) the original RGB visualisation of the FT-IR image, (b) freeware software siMPle \cite{primpke2020toward} marching results (c) and the semi-supervised method with clustering + cluster centroid spectral matching. For the image evaluated here, the proposed workflow required approximately 2~min for clustering and 0.23~min for spectral matching, whereas processing with siMPle required approximately 1.5~h for 12 reference spectra. These values provide an empirical comparison for the analysed image and computational environment; they should not be interpreted as a benchmark of scaling with image resolution.}
    \label{fig:theMaps}
\end{figure}

A comparison between the proposed clustering-based approach and the freeware plastic identification software siMPle \cite{primpke2020toward} is presented in Fig. \ref{fig:theMaps}. While both methods achieve comparable identification at a coarse level, the clustering-based strategy yields particle maps that more accurately preserve particle morphology. In particular, the spatial extent and boundaries of individual particles appear more coherent and less fragmented compared to the siMPle output.

This improved shape fidelity is attributed to the use of clustering prior to spectral matching, which enforces spatial and spectral consistency within particle regions. As a result, the method reduces pixel-level misclassification and noise, leading to more contiguous particle representations. This is especially beneficial for downstream tasks such as particle counting, where accurate delineation of particle boundaries is critical.

Overall, the results in Fig. \ref{fig:theMaps} suggest that incorporating clustering as an intermediate step enhances not only classification robustness but also the spatial interpretability of the resulting particle maps.

The comparison of spectral matching strategies in Table \ref{tab:matching_accuracy} indicates that the sign-invariant derivative-based method (M5) provides the most suitable pairing with cluster centroid matching, achieving perfect classification accuracy across all clusters. In contrast, most alternative methods exhibit substantially lower performance, with several approaches (M2–M4, M7, M9) failing to correctly classify any clusters under the same conditions.

The superior performance of the derivative-based M5 can be attributed to its reliance on derivative spectral features combined with sign-invariant cosine similarity, which emphasizes spectral shape while mitigating sensitivity to baseline variations and sign ambiguities. This is particularly advantageous in the present framework, where cluster centroids represent averaged spectra that may differ in scaling or orientation from reference spectra.

The cluster-level similarity scores for M5 further illustrate consistent and robust matching across all classes, with values ranging from 0.603 to 0.836. The highest score is observed for Cluster 3 (0.836), indicating strong agreement with the corresponding reference class, while the remaining clusters also achieve comparatively high similarity values (C1: 0.694, C2: 0.603, C4: 0.663). Despite moderate variation in absolute similarity, all clusters are correctly classified, demonstrating that the method provides reliable discrimination even when spectral agreement is not uniformly maximal.

Overall, these results highlight that derivative-based, sign-invariant similarity measures are particularly well aligned with cluster-based representations, enabling robust and consistent identification across all evaluated polymer classes.

\begin{table}[H]
\centering
\small
\setlength{\tabcolsep}{6pt}
\renewcommand{\arraystretch}{1.1}

\begin{tabular}{ccc}
\hline
\textbf{Method} & \textbf{Matching accuracy} & \textbf{M5 cluster scores} \\
\hline
M1  & 0.25 &  \\
M2  & 0.00 & \\
M3  & 0.00 &  \\
M4  & 0.00 &  \\
M5  & \textbf{1.00} & \begin{tabular}[c]{@{}c@{}}C1: 0.694 \\ C2: 0.603 \\ C3: 0.836 \\ C4: 0.663\end{tabular} \\
M6  & 0.50 &  \\
M7  & 0.00 & \\
M8  & 0.25 & \\
M9  & 0.00 & \\
M10 & 0.75 & \\
M11 & 0.75 & \\
M12 & 0.25 &  \\
\hline
\end{tabular}

\caption{Matching accuracy of the evaluated spectral-library matching methods. The sign-invariant derivative-based method (M5) achieved perfect classification performance.}
\label{tab:matching_accuracy}
\end{table}

\begin{figure}[H]
    \centering
    \begin{subfigure}[b]{0.48\linewidth}
        \centering
        \includegraphics[width=0.9\linewidth]{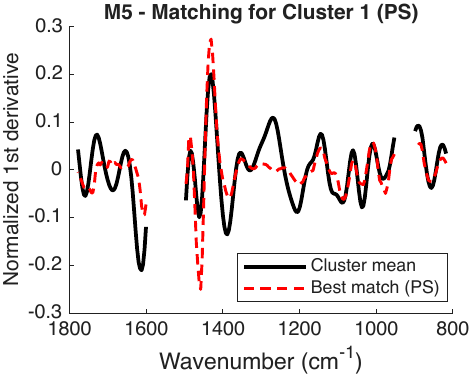}
        \caption{}
        \label{fig:PSmatch}
    \end{subfigure}
    \begin{subfigure}[b]{0.48\linewidth}
        \centering
        \includegraphics[width=0.9\linewidth]{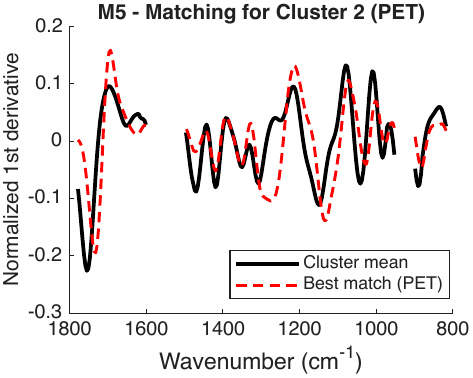}
        \caption{}
        \label{fig:PETmatch}
    \end{subfigure}
    \begin{subfigure}[b]{0.48\linewidth}
        \centering
        \includegraphics[width=0.9\linewidth]{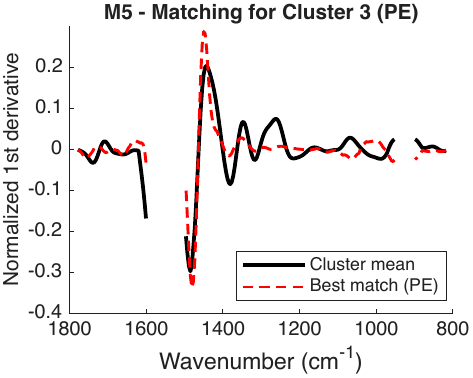}
        \caption{}
        \label{fig:PEmathc}
    \end{subfigure}
    \begin{subfigure}[b]{0.48\linewidth}
        \centering
        \includegraphics[width=0.9\linewidth]{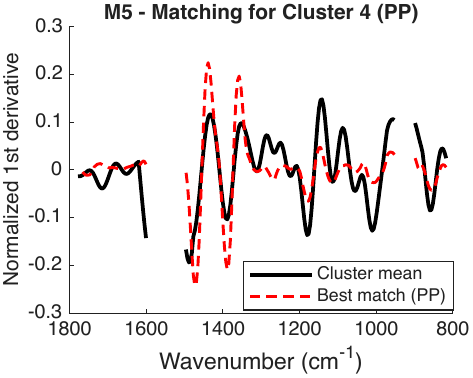}
        \caption{}
        \label{fig:PPmatch}
    \end{subfigure}
    \caption{The cluster centroids matches along with the reference spectra for each matched MP type for (a) PS, (b) PET, (c) PE, (d) PP.}
    \label{fig:theMatchings}
\end{figure}

Figure~ \ref{fig:theMatchings} presents the cluster centroid spectra alongside their corresponding best-matching reference spectra for each polymer class. Overall, the centroid spectra show good agreement with the reference profiles, capturing the main characteristic absorption bands that define each material.

However, systematic differences between sample and reference spectra can also be observed. In particular, additional peaks appear in the measured microplastic spectra that are not present, or are less pronounced, in the reference library. A notable example is the emergence of a band in the \SIrange{1250}{1300}{\per\centi\meter} region across all polymer types. Such features may arise from environmental effects, including weathering, surface contamination, additives, or differences in sample morphology and thickness, all of which are known to alter spectral signatures in real-world microplastic samples.

Although some differences were observed between the measured spectra and the reference spectra, the main spectral features were still similar enough to allow correct classification. This suggests that the selected matching method (M5) is useful because it compares the overall shape of the spectra, rather than depending only on exact peak positions or intensities. As a result, it can tolerate moderate spectral differences while still identifying the polymer type correctly.

\subsection{Supervised classification}

For the supervised calibration, the validation and test results presented are for the public datasets with labeled pixel-data. Then, the in-house inquired data for the best performing models with the publicly-available datasets, are visually evaluated. The summary of the results is presented in Table~\ref{tab:best_supervised_configs}. From the results presented, the following observations can be drawn:
\begin{itemize}
    \item Test accuracy and weighted F1-score were nearly identical for the best-performing models, indicating that high accuracy was not driven only by majority-class performance.
    \item The strongest configurations most often involved derivative- or smoothing-based representations, suggesting that spectral-shape information was more informative than raw intensity alone.
    \item Tree ensembles, KNN models, and shallow neural networks were the most consistently competitive model families, whereas Naive Bayes and high-order SVM kernels were more sensitive to preprocessing choices.
    \item Although several models achieved very high test performance on public labelled datasets, this should not be interpreted as proof of transferability to in-house FT-IR images. The in-house maps are therefore used as an external qualitative check of spatial coherence and particle-level plausibility.
    \item The large spread in performance across preprocessing configurations highlights that preprocessing should be treated as part of the model selection problem, rather than as a fixed preliminary step.
\end{itemize}

\begin{table}[H]
\centering
\scriptsize
\setlength{\tabcolsep}{3pt}
\renewcommand{\arraystretch}{1.1}
\caption{Best preprocessing configuration obtained for each supervised classification method. Each model has been trained on a full-factorial of 96 pre-treatment configurations, and the results for the best configuration has been reported. The reported metrics are class-weighted accuracy (WAcc.) and class-weighted F1-score (WF1).}
\label{tab:best_supervised_configs}
\begin{tabularx}{\linewidth}{p{0.15\linewidth} p{0.24\linewidth} X c c}
\hline
\textbf{Model type} & \textbf{Model} & \textbf{Best configuration (out of 96)} & \textbf{Test WAcc. (\%)} & \textbf{Test WF1 (\%)} \\
\hline
Ensemble & Bagged Trees & SG(2,7) + first derivative + row-centering  & 99.832 & 99.832 \\
Ensemble & RUSBoosted Trees & AsLS + SG(2,7) + first derivative + row-centering & 99.579 & 99.579 \\
Neural Network & Bilayered Neural Network & SG(2,11) smoothing + variable-wise z-score  & 99.579 & 99.579 \\
KNN & Fine KNN &  SG(2,7) + first derivative + row-centering  & 99.495 & 99.495 \\
KNN & Weighted KNN & SG(2,7) + first derivative + row-centering  & 99.495 & 99.495 \\
Ensemble & Subspace Discriminant & SG(2,7) + AsLS + SNV  & 99.495 & 99.493 \\
Ensemble & Subspace KNN & SG(2,7) + first derivative + variable-wise z-score & 99.495 & 99.495 \\
Neural Network & Narrow Neural Network & SG(2,15) + variable-wise z-score & 99.495 & 99.495 \\
Neural Network & Medium Neural Network & SG(2,11) + variable-wise z-score & 99.495 & 99.495 \\
Neural Network & Wide Neural Network & SG(2,7) smoothing & 99.495 & 99.495 \\
SVM & Quadratic SVM & AsLS +  SG(2,7) + SNV  & 99.411 & 99.412 \\
Ensemble & Boosted Trees & SG(2,7) smoothing + SNV  & 99.411 & 99.411 \\
Neural Network & Trilayered Neural Network & SG(2,11) + Numerical gradient & 99.411 & 99.411 \\
KNN & Cosine KNN & SG(2,7) + First finite difference  & 99.243 & 99.243 \\
PLS & PLS-DA & SG(2,11) smoothing + variable-wise z-score  & 99.092 & 99.092 \\
SVM & Medium Gaussian SVM & SG(2,7) + SNV  & 98.991 & 98.989 \\
KNN & Cubic KNN &  SG(2,7) + Numerical gradient & 98.907 & 98.903 \\
Tree & Fine Tree & AsLS + SG(2,15) + first derivative  & 98.823 & 98.825 \\
SVM & Linear SVM & SG(2,11) smoothing + SNV & 98.823 & 98.821 \\
KNN & Medium KNN & AsLS + SG(2,7) first derivative + row-centering & 98.823 & 98.823 \\
Discriminant & Linear Discriminant & AsLS + SG(2,7) + SNV & 98.738 & 98.738 \\
Tree & Medium Tree & SG(2,7) + Numerical gradient + row-centering & 98.654 & 98.652 \\
Efficient Linear & Efficient Linear SVM & SG(2,11) + first derivative + SNV & 98.402 & 98.404 \\
Efficient Linear & Efficient Logistic Regression & AsLS + SG(2,11) + SNV & 97.225 & 97.224 \\
SIMCA & SIMCA & SG(2,11) smoothing + variable-wise z-score  & 95.632 & 95.502 \\
KNN & Coarse KNN & AsLS + SG(2,7) + first derivative & 94.617 & 94.493 \\
SVM & Fine Gaussian SVM & SG(2,15) smoothing + SNV & 94.281 & 94.215 \\
Tree & Coarse Tree & SG(2,7) + First finite difference + SNV  & 93.776 & 93.705 \\
SVM & Coarse Gaussian SVM & SG(2,7) + first derivative + SNV & 93.776 & 93.615 \\
Naive Bayes & Kernel Naive Bayes & SG(2,11) + first derivative + SNV & 92.178 & 92.040 \\
Naive Bayes & Gaussian Naive Bayes & SG(2,7) + first derivative + SNV & 91.001 & 90.974 \\
SVM & Cubic SVM & SG(2,7) + first derivative + SNV  & 77.376 & 72.310 \\
\hline
\end{tabularx}
\end{table}

\begin{figure}[H]
    \centering
    \begin{subfigure}[b]{0.32\linewidth}
        \includegraphics[width=\linewidth]{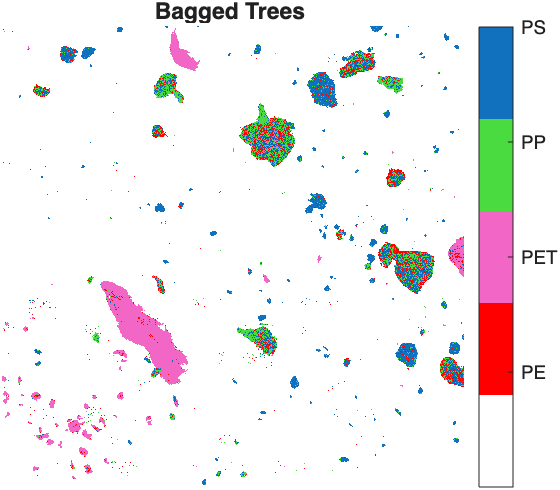}
        \caption{}
        \label{fig:method1SC}
    \end{subfigure}
    \begin{subfigure}[b]{0.32\linewidth}
        \includegraphics[width=\linewidth]{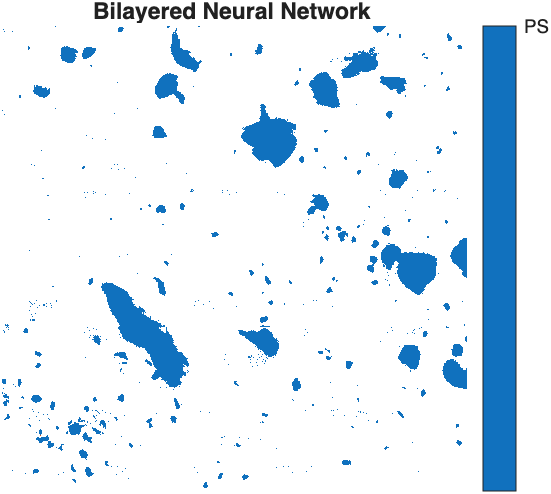}
        \caption{}
        \label{fig:method2SC}
    \end{subfigure}
    \begin{subfigure}[b]{0.32\linewidth}
        \includegraphics[width=\linewidth]{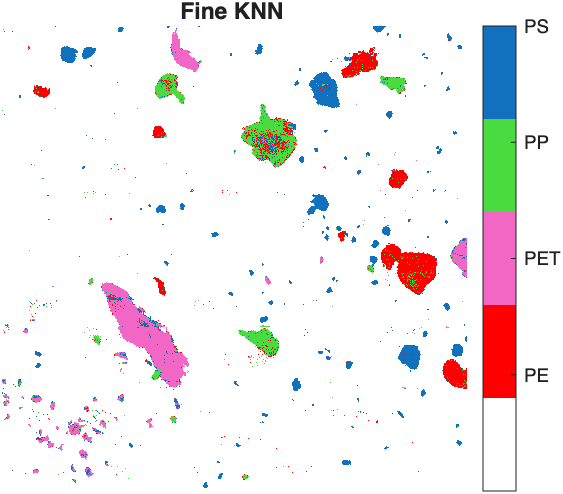}
        \caption{}
        \label{fig:method3SC}
    \end{subfigure}
    \caption{The classification maps of the best-performing methodologies, for the in-house acquired image: (a) bagged trees (b) bilayered neural network and (c) fine KNN.}
    \label{fig:supervisedMethodsVisu}
\end{figure}

However, despite their very good performance with the three public datasets, the performance is not transferred to our in-house acquired data. Figure~\ref{fig:supervisedMethodsVisu} showcases supervised classification results on the in-house acquired image. Even though the illustrated methods had very high classification performance on the three public datasets data (over 99\%), their performance on the test dataset is significantly lower that the unsupervised clustering and spectral matching approach. The following issues can be identified for the bagged trees and the k-nearest neighbor methods: (a) some particles are completely miss-identified, with pixels inside them belonging to all four MP categories; (b) a heavy PP-PE miss-classification can be observed. The bilayered neural network, even though it achieved almost perfect classification with the three public datasets, classified all the particle pixels to the PS type. This indicates great over-fitting to the training datasets, and very poor model transferability. 

The pronounced difference between the results obtained for the public datasets and the in-house dataset indicates a substantial dataset-shift problem. Several factors may contribute simultaneously to this discrepancy. These include differences in spectral resolution and instrumental response, a lower signal-to-noise ratio in experimentally acquired particle spectra, differences in particle-size and morphology distributions, and chemical or baseline offsets caused by environmental weathering, surface contamination, or sample preparation. Because these factors were not independently controlled and vary concurrently among the datasets, their individual contributions cannot be isolated quantitatively in the present study. Accordingly, the results should not be interpreted as establishing a single cause for the performance decrease. Instead, they demonstrate that near-perfect classification within curated public datasets does not necessarily indicate reliable transfer to independently acquired and more heterogeneous samples.

The known-polymer samples were pristine materials whose identities were based on the information supplied by the producer and were independently confirmed using ATR-FTIR. Consequently, the interpretation of these results confirms that the producer-provided polymer identities were correct. Independent confirmation using ATR-FTIR provides an additional level of validation.

\section{Conclusions}

This study proposed a multi-level preprocessing and modelling framework for FT-IR spectral imaging of microplastics, combining image-level, tile-level, and spectral-level corrections with scalable polymer identification. The results show that acquisition-related variability can be identified and reduced using background-referenced PCA, while tile-repeated artefacts can be mitigated through background-based artefact estimation.  These correction steps improved the spatial coherence of the resulting particle masks and reduced the apparent fragmentation of individual particles into multiple disconnected regions, as illustrated by the comparison with siMPle.

The proposed semi-supervised workflow, based on particle masking, clustering, and spectral matching of cluster centroids, provided robust and computationally efficient microplastic identification. Compared with direct pixel-wise software-based matching, the clustering-based approach produced more spatially coherent particle maps and substantially reduced processing time. Among the twelve evaluated matching strategies, the sign-invariant derivative-based cosine similarity method achieved the best performance, correctly identifying all evaluated polymer classes: PS, PET, PE, and PP. 

The supervised classification results further demonstrated that high performance on public labelled datasets does not necessarily guarantee transferability to independently acquired FT-IR images. Although several models achieved weighted accuracy and F1-scores above 99\% on public datasets, their predictions on the in-house image were less reliable, with visible particle-level misclassification and strong PE--PP confusion. This highlights the risk of relying only on aggregate validation metrics and emphasizes the need for external image-level evaluation when developing microplastic identification models.

Overall, the results suggest that robust microplastic spectral-imaging workflows should treat preprocessing, artefact correction, particle segmentation, clustering, and identification as interconnected steps rather than independent procedures. The proposed framework improves robustness, interpretability, and computational scalability by reducing acquisition artefacts, retaining only particle spectra, and performing library matching on representative cluster centroids. 

The results demonstrate that the proposed particle-pixel-wise workflow can provide a substantial reduction in processing time for the images evaluated in this study. However, computational scaling with image resolution was not systematically investigated. In particular, the present results do not establish linear scaling when the dimensions of the input image are increased. Such scaling is expected to depend not only on the total number of image pixels, but also on the proportion of pixels retained as candidate particles, the number and morphology of the particles, the clustering configuration, and the computational hardware. Future benchmarking should therefore vary both image dimensions and particle coverage and report the computational cost of each processing stage separately.

Future work should extend the framework to more polymer types, weathered and environmentally contaminated particles, and larger multi-instrument datasets, as well as incorporate uncertainty-aware decision rules for highly confusable polymer classes. The construction of comprehensive and representative reference libraries remains a major challenge in microplastic spectroscopy. A library intended for robust practical identification should represent not only different polymer classes, but also variability arising from instruments and acquisition settings, spectral resolution, signal quality, particle size and morphology, environmental ageing, surface contamination, and sample preparation. Future work should evaluate these factors through controlled experiments in which spectral resolution, noise level, particle size, and weathering state are varied independently. This would enable their individual and interacting effects on model transferability to be quantified and could support the development of more reliable cross-dataset calibration and classification strategies.

\section*{Acknowledgements}

ZSD, SH, TS, and SPR acknowledge funding from Research Council of Finland for the Flagship of Advanced Mathematics for Sensing, Imaging, and Modelling 2024--2031 (decision number 359183). 

\section*{Ethics declarations}
\textbf{Ethical approval} \\
Not applicable.

\textbf{Competing interests} \\
The authors declare no competing interests. 

\appendix

\section{Time-dependent interference analysis}\label{app:tdi}

\begin{figure}[H]
    \centering

    \begin{subfigure}[b]{0.99\linewidth}
        \centering
        \includegraphics[width=\linewidth]{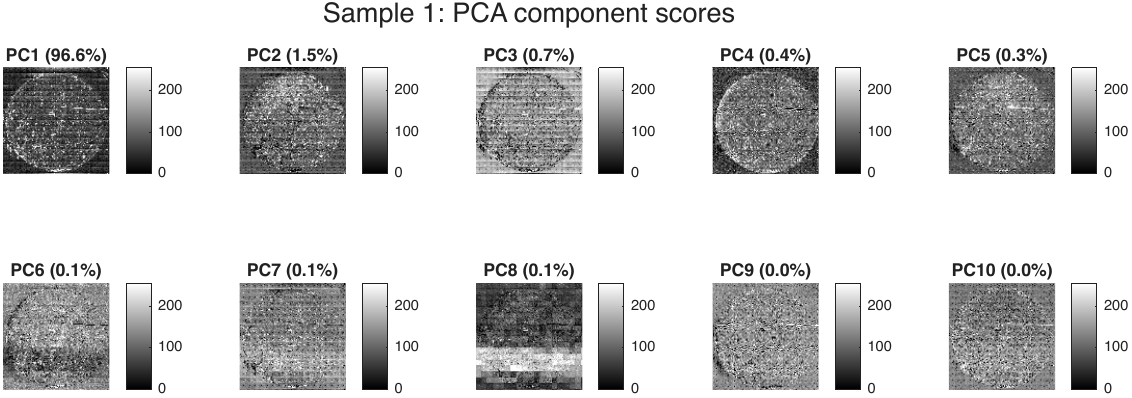}
        \caption{Sample~1, exhibiting temporally structured variation during the second half of the tile-by-tile acquisition.}
        \label{fig:sample1}
    \end{subfigure}
    \hfill
    \begin{subfigure}[b]{0.99\linewidth}
        \centering
        \includegraphics[width=\linewidth]{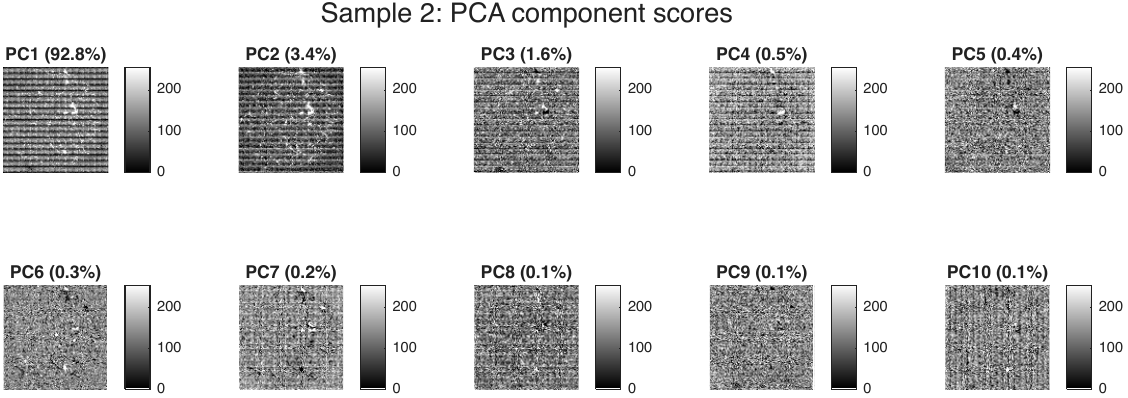}
        \caption{Sample~2, exhibiting no comparable acquisition-dependent pattern among the inspected principal components.}
        \label{fig:sample2}
    \end{subfigure}

    \caption{Grayscale representations of the principal-component score maps for the two independently analysed samples. Identically numbered PCs in the two panels are not directly equivalent because PCA was fitted independently to each sample.}
    \label{fig:grayscalescores}
\end{figure}

To illustrate the identification of temporally structured background variation, two samples acquired using the same experimental workflow were compared. The samples are different than the sample illustrated in the main body of the article, but contain the same microplastics types. Sample~1 exhibited a systematic change during the second half of the tile-by-tile acquisition, whereas Sample~2 showed no comparable acquisition-dependent pattern. Figure~\ref{fig:grayscalescores} presents the principal-component score maps converted to grayscale values in the range 0--255. For each PC, the score values were clipped to the 2nd--98th percentile interval and linearly rescaled to facilitate visualization. The same grayscale images were subsequently used to extract the background-ROI distributions and calculate the Wasserstein distances. In Sample~1, temporally structured tile patterns are visually apparent in PCs~6--8, which together account for approximately 0.3\% of the total variance in the mean-centred PCA model fitted to the reduced $300\times300$-pixel image. In contrast, no comparable pattern is apparent among the inspected PCs of Sample~2.

\begin{figure}[H]
    \centering

    \begin{subfigure}[b]{0.99\linewidth}
        \centering
        \includegraphics[width=\linewidth]{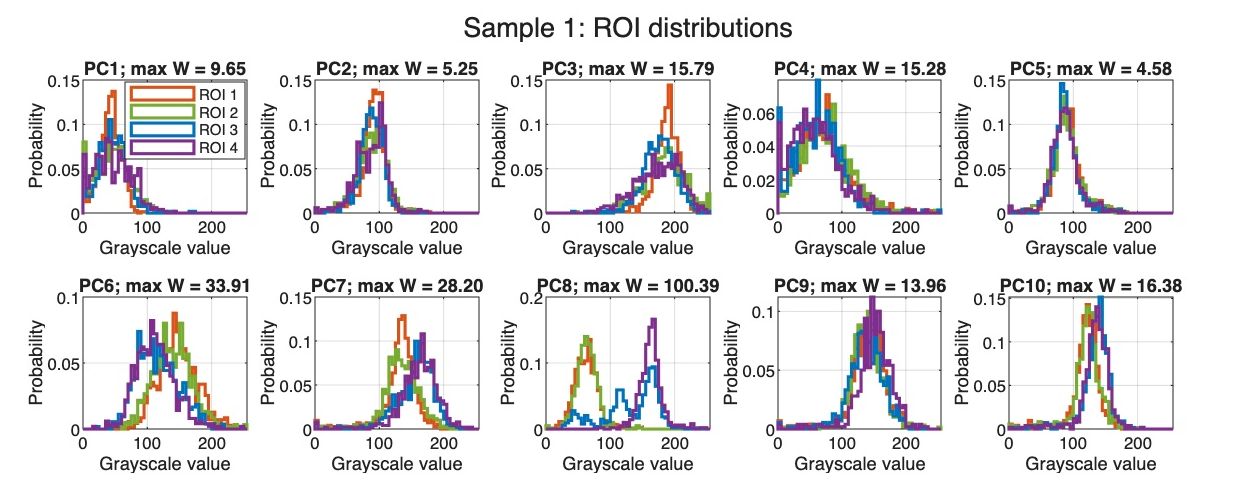}
        \caption{}
        \label{fig:Wdist1}
    \end{subfigure}
    \hfill
    \begin{subfigure}[b]{0.99\linewidth}
        \centering
        \includegraphics[width=\linewidth]{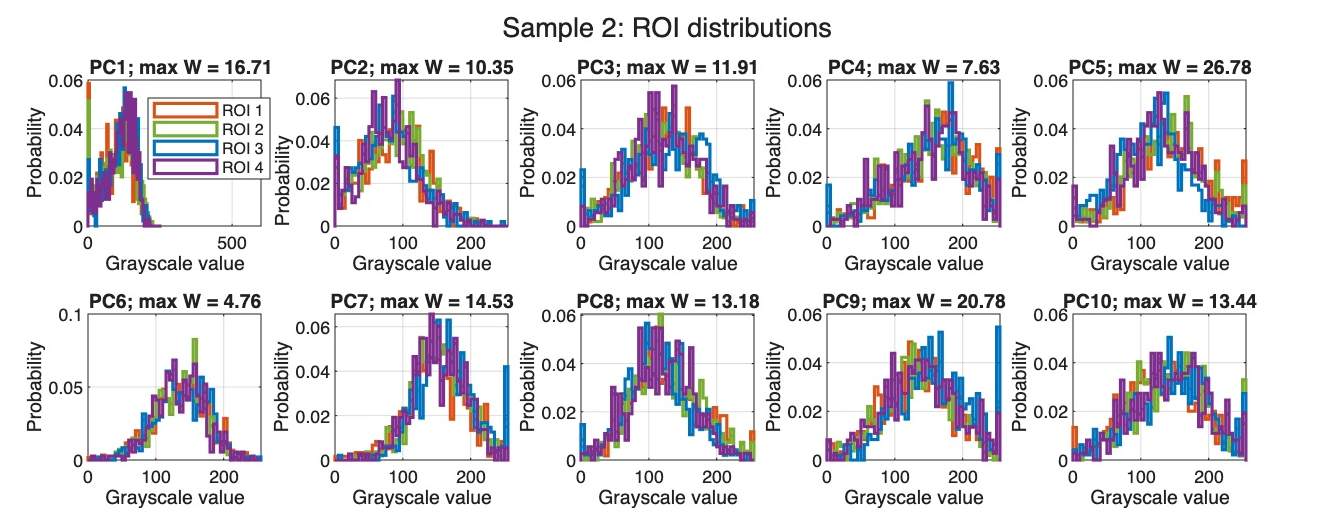}
        \caption{}
        \label{fig:Wdist2}
    \end{subfigure}
    \hfill
    \begin{subfigure}[b]{0.99\linewidth}
        \centering
        \includegraphics[width=\linewidth]{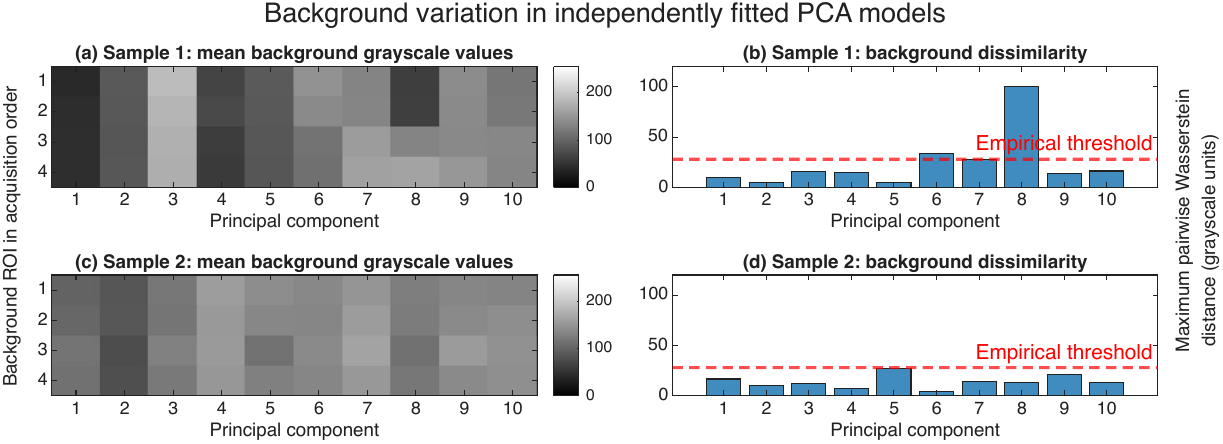}
        \caption{}
    \end{subfigure}
    \caption{Background-referenced assessment of acquisition-dependent variation in the independently fitted PCA models. For each principal component, grayscale-value distributions were extracted from four fixed, particle-free background ROIs, and all six pairwise first-order Wasserstein distances were calculated. The maximum pairwise distance for each PC was compared with an empirical threshold of 30 grayscale units.}
    \label{fig:wasdistances}
\end{figure}

The visual assessment is supported by the Wasserstein-distance analysis presented in Fig.~\ref{fig:wasdistances}. Pronounced differences among the background-ROI grayscale distributions are observed for PCs~6--8 of Sample~1. For the rescaled 0--255 grayscale images, an empirical Wasserstein-distance threshold of 30 grayscale units was applied. Based on this criterion, PCs~6 and 8 of Sample~1 exceeded the threshold and were identified as nuisance components. Although a structured pattern was also visually apparent in PC~7, its maximum pairwise Wasserstein distance remained below the empirical threshold. For Sample~2, the pairwise distributional differences remained within the empirical limit for all inspected PCs, and no nuisance components were identified.

%\bibliography{bibliography}

\end{document}